\documentclass[twocolumn]{aastex631}

\usepackage{ gensymb }
\usepackage{textcomp}
\usepackage{mathtools}
\usepackage{booktabs}
\usepackage{amsmath}	
\usepackage{comment}

\accepted{March 09, 2024}
\submitjournal{ApJL}

\shorttitle{Overdense filamentary structure at $z \sim$1.5}
\shortauthors{Laishram et al.}

\begin{document}

\title{Insights into Galaxy Morphology and Star Formation: Unveiling Filamentary Structures around an Extreme Overdensity at $z \sim 1.5$ Traced by [OII] Emitters}

\correspondingauthor{Ronaldo Laishram}
\email{ronaldolaishram@astr.tohoku.ac.jp}

\author [0000-0002-0322-6131]{Ronaldo Laishram}
\affiliation{Astronomical Institute, Tohoku University, 6-3, Aramaki, Aoba, Sendai, Miyagi, 980-8578, Japan}
\affiliation{Graduate Program on Physics for the Universe (GP-PU), Tohoku University, 6–3 Aoba, Sendai 980-8578, Japan}
\author [0000-0002-2993-1576]{Tadayuki Kodama}
\affiliation{Astronomical Institute, Tohoku University, 6-3, Aramaki, Aoba, Sendai, Miyagi, 980-8578, Japan}

\author [0000-0002-8512-1404]{Takahiro Morishita}
\affiliation{IPAC, California Institute of Technology, MC 314-6, 1200 E. California Boulevard, Pasadena, CA 91125, USA}

\author [0000-0002-9382-9832]{Andreas Faisst}
\affiliation{IPAC, California Institute of Technology, MC 314-6, 1200 E. California Boulevard, Pasadena, CA 91125, USA}

\author [0000-0002-0479-3699]{Yusei Koyama}
\affiliation{Subaru Telescope, National Astronomical Observatory of Japan, 650 North A’ohoku Place, Hilo, HI 96720, USA}

\author {Naoaki Yamamoto}
\affiliation{Astronomical Institute, Tohoku University, 6-3, Aramaki, Aoba, Sendai, Miyagi, 980-8578, Japan}
\affiliation{Graduate Program on Physics for the Universe (GP-PU), Tohoku University, 6–3 Aoba, Sendai 980-8578, Japan}

\begin{abstract}
We explore the morphological features and star formation activities of [OII] emitters in the COSMOS UltraDeep field at $z \sim 1.5$ using JWST NIRCam data from the COSMOS-Web survey and Subaru Hyper Suprime-Cam. We also report the discovery of large filamentary structures traced by [OII] emitters, surrounding an extremely overdense core with a galaxy number density $\sim11\times$ higher than the field average. These structures span over 50 cMpc, underscoring their large scale in the cosmic web at this epoch. After matching the stellar mass distributions, the core galaxies show a higher frequency of disturbances (50\% $\pm$ 9\%) than those in outskirts (41\% $\pm$ 9\%) and the field (21\% $\pm$ 5\%), indicative of more frequent mergers and interactions in the innermost $\lesssim1.5$\,arcmin region.
Additionally, we observe that specific star formation rates are elevated in denser environments. A Kolmogorov-Smirnov (KS) test comparing the distribution of specific star formation rates of core and field galaxies yields a \textit{p}-value of 0.02, suggesting an enhancement of star-formation activity driven by the dense environment. Our findings underscore the environmental impact on galaxy evolution during a pivotal cosmic epoch and set the stage for further investigation with the increasing larger data from upcoming surveys.
\end{abstract}

\keywords{Galaxy evolution (594); Galaxy environments (2029); Galaxy structure (622); Galaxy properties (615)}

\section{\textbf{Introduction}} \label{sec:intro}

The environment of galaxies is known to play a significant role in shaping galaxy evolution by influencing factors such as morphology, gas content, and star formation rates. 
In the local universe galaxy properties vary noticeably along the cosmic web, from dense clusters to more isolated fields. Passive galaxies dominate dense regions, whereas star-forming galaxies are more prevalent in field environments \citep[e.g.,][]{1980ApJ...236..351D, 1998ApJ...499..589H, 1999ApJ...518..576P, 2001ApJ...549..820C, 2003MNRAS.346..601G, 2004MNRAS.353..713K,2004ApJ...615L.101B, 2007ApJS..172..284C,morishita17}.

The influence of environment on the properties of galaxies has been observationally revealed across various redshift ranges. Parameters such as the star formation rate (SFR), galaxy morphology, and stellar mass are strongly influenced by the galaxy number density in their surrounding universe \citep{1980ApJ...236..351D, 2001MNRAS.326..637K, 2001ApJ...562L...9K, 2010MNRAS.408.1417S, 2010ApJ...721..193P}. 
For instance, galaxies in lower-density environments exhibit higher median specific star formation rates (sSFR) at lower redshifts, indicating that environmental factors play a significant role in determining the fraction of quiescent galaxies. Further studies have demonstrated that this relationship evolves with redshift, with a notable reversal occurring around z $\ge$1, where star formation activity begins to correlate positively with the density of galaxies \citep{2007A&A...468...33E, 2008MNRAS.383.1058C, 2022A&A...662A..33L}. It was also found that blue-fraction of galaxies in clusters increased with redshifts \citep{1978ApJ...219...18B, 1984ApJ...285..426B}. This evolution is intricately linked to the morphology–density relationship, where dense environments typically host quiescent, bulge-dominated galaxies, contrasting with the more active, star-forming galaxies in lower density environments \citep{1980ApJ...236..351D}. This relationship is a pivotal aspect in understanding galaxy evolution, as it highlights the influence of the local environment on galaxy morphology and star formation activities.

The morphology-density relationship at higher redshifts remains a question. Recent studies extend this understanding to higher redshifts, exploring the morphology–density relationship in distant galaxy clusters \citep{2014ApJ...788...51N, 2019A&A...630A..57P}. \citet{2009A&A...503..379T} found that the morphology- density relation is already in place at $z =$ 1, but the trend is flatter towards higher redshifts, indicating the morphological evolution of cluster galaxies since $z =$ 1. \citet{2020ApJ...899...85S} found that the environmental processes influencing galaxy shapes and densities were active as early as $z =$1.75, primarily affecting low-mass galaxies. \citet{2023A&A...670A..58M} found that the fraction of spheroid galaxies is higher in clusters, as well as the fraction of mergers. 
Star-forming galaxies in a $z\sim$2 protocluster exhibit morphologies similar to those in the general field, as reported in several studies \citep{2007ApJ...668...23P, 2023ApJ...958..170N}.

Numerical studies have enabled detailed morphological predictions \citep{2015MNRAS.446..521S, 2019MNRAS.490.3234N}, at even the highest redshifts \citet{2022MNRAS.514.1921R}. However, reconciling these predictions with observational data presents challenges, especially in light of the James Webb Space Telescope (JWST) observations that show a diversity of galaxy structures at higher redshifts \citep{2022ApJ...938L...2F, 2023ApJ...955...94F,2023ApJ...946L..15K}. This discrepancy underscores the need for a more thorough understanding of morphological evolution and its interplay with environmental factors at high redshifts. The JWST has revolutionized our view of the distant universe, offering unprecedented insights into galaxy assembly during this transformative epoch and allowing us to test key theoretical predictions about the early universe.

In this Letter, we report the discovery of an overdensity filamentary structures traced by [OII] emitters at $z \sim$1.5 in the COSMOS-UltraDeep field. By examining the morphologies of galaxies using recently acquired JWST NIRCam data, we investigate the impact of various environments on their ongoing star formation activities.
Throughout this paper, the concordance $\Lambda$CDM cosmology (H$_0$ = 70 km s$^{-1}$ Mpc$^{-1}$ 
, $\Omega_{\Lambda}$ = 0.7 and $\Omega_{M}$ = 0.3) is adopted, and all magnitudes are reported on the AB system \citep{1983ApJ...266..713O}. Finally, \cite{Chabrier_2003} initial mass function (IMF) is adopted.

\begin{figure*}
\centering
\includegraphics[width=0.75\linewidth]{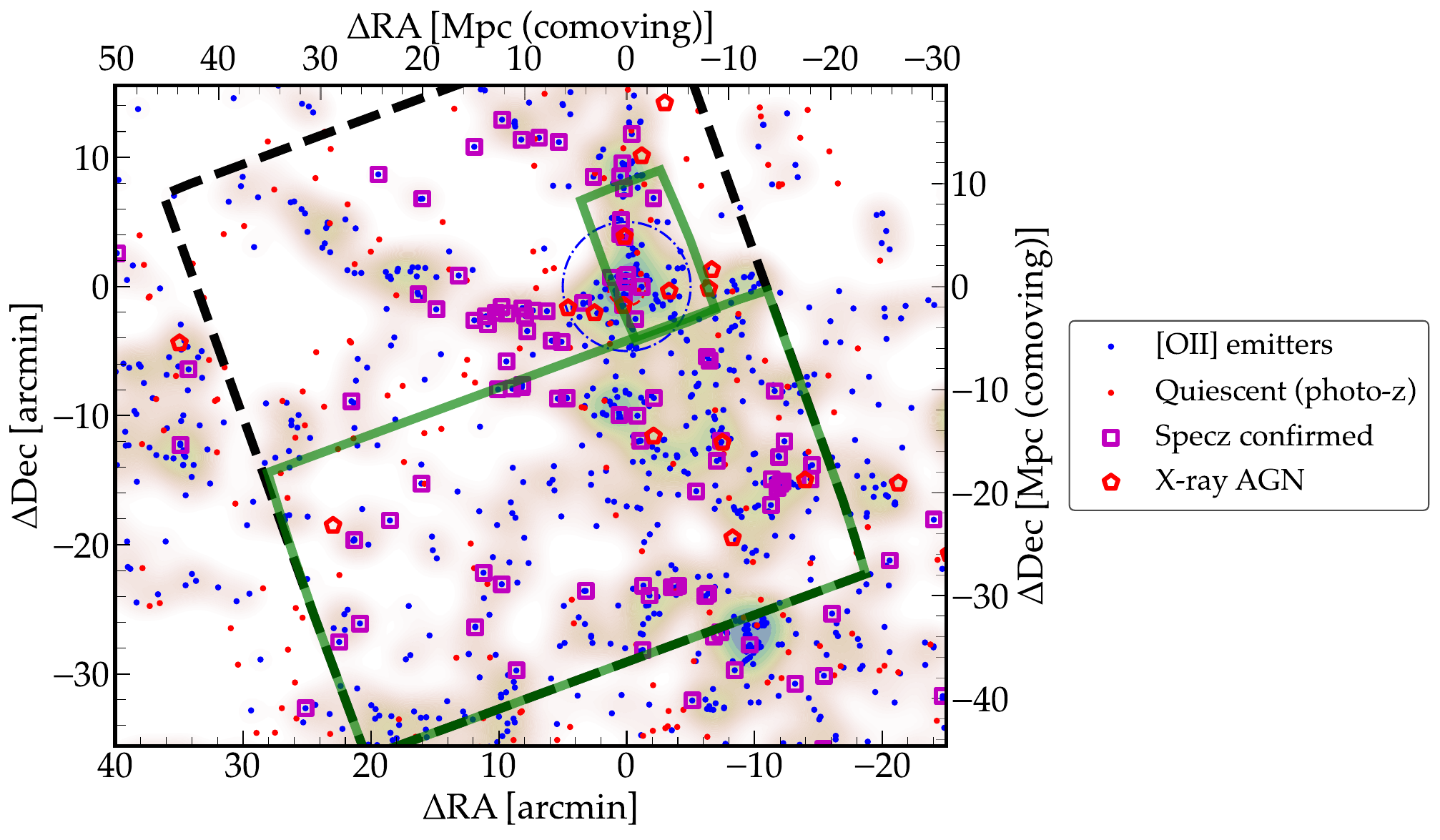} %
\vspace{4pt} 
\includegraphics[width=0.95\linewidth]{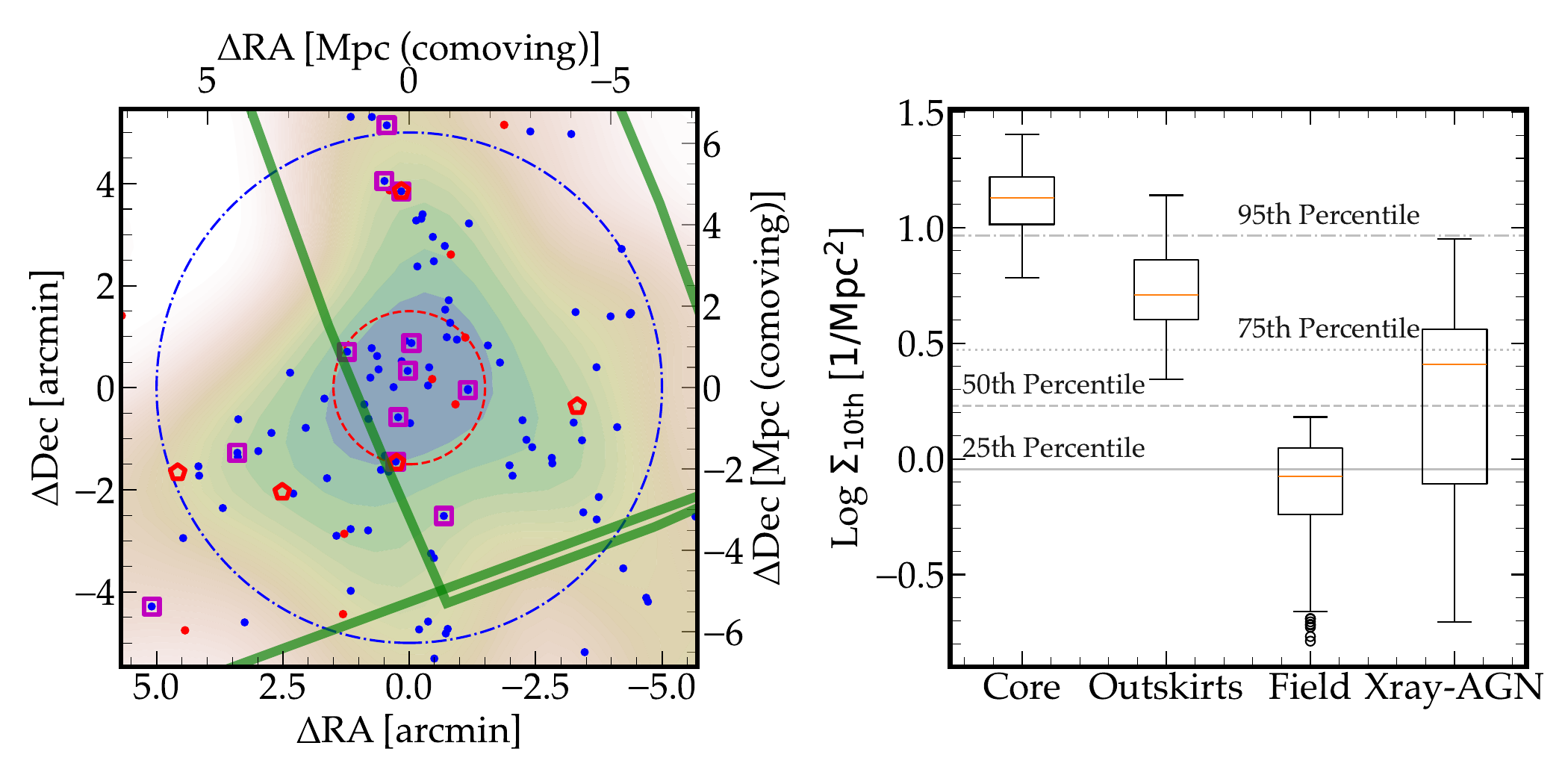} %
\caption{Top: Spatial distribution of [OII] emitters identified in our narrow-band selection (blue points) and various populations indicated by distinct colors and markers: Quiescent galaxies (red points), X-ray AGN (red pentagons). Spec-$z$ confirmed sources are marked by magenta squares. The full footprint of COSMOS-Web \citep{2023ApJ...954...31C} is outlined with black dashed line, whereas the green line represents the areas this study investigates (Epoch 1: January 5/6, 2023; Epoch 2: April/May 2023). Density distribution visualized using a kernel density estimate (KDE) with a bandwidth adjustment factor of 0.2, overlaid on contour levels. Bottom (left): close-up view of the core ($r \leq 1.5'$; red line) and outskirt regions ($1.5' < r \leq 5'$; blue). Bottom (right): Boxplot comparison of the local density $\Sigma_{10\mathrm{th}}$ [1/Mpc$^2$] across core, outskirts, field environments, and X-ray detected AGNs. The dashed horizontal lines indicate the 25th, 50th, 75th, and 95th percentiles of the combined density distribution.}
\label{fig:cosmic_structure}
\end{figure*}

\section{\textbf{DATA}}\label{sec:data}

\subsection{[OII] emitter samples}\label{sec:sample}
Our study focuses on the [OII] emitters selected by a narrow-band (NB) filter, NB921, centered at $z\sim1.471$ (ranging from 1.453–1.489) within the COSMOS UltraDeep (COSMOS-UD) field \citep{2020PASJ...72...86H}. COSMOS-UD is one of the surveys of the Hyper Suprime-Cam Subaru Strategic Program \citep{2018PASJ...70S...4A}. We employed the NB921 filter data from the Hyper Suprime-Cam Subaru Strategic Program's second public data release (PDR2) \citep{2019PASJ...71..114A}. [OII] emitters were identified using the narrow band imaging, a technique effective for isolating star-forming galaxies with specific nebular emission lines at a particular redshift. This method ensures a homogeneous field-of-view and minimizes bias in target selection. Additionally, due to the narrow redshift ranges surveyed by NB imaging, the distribution of emission-line galaxies selected is minimally affected by contamination from fore- and background sources, making it an ideal approach for investigating the properties of star-forming galaxies and their environmental dependencies.  For a detailed description of the emission-line galaxy selection process and the methods applied to this data set, we refer to \citet{2020PASJ...72...86H} for an in-depth overview of the selection procedure.

Our sample consists of 1449 [OII] emitters over the survey area of 1.37 $\mathrm{deg^2}$. To assess the purity of this sample and confirm the [OII] emission status, we cross-matched our candidates with existing spectroscopic redshifts from various surveys conducted in the COSMOS field, including the Keck/DEIMOS survey \citet{2018ApJ...858...77H}, the Fiber Multi-Object Spectrograph (FMOS)-COSMOS program \citep{2017ApJ...835...88K, 2015ApJS..220...12S} and MOSFIRE Deep Evolution Field (MOSDEF) survey \citep{2015ApJS..218...15K}. This comparison yielded 115 emitters with corresponding spectroscopic redshifts (spec-$z$), of which 113 were confirmed to have spec-$z$ values at or near $z\sim $1.5, aligning with the expected redshift for [OII] emitters. The remaining two sources were identified at a lower redshift, suggesting minimal contamination within our narrow-band selection. Subsequently, these two lower-redshift sources are removed from further analysis.

We supplement our samples with photometric sources selected from the COSMOS2020 CLASSIC catalog \citep{2022ApJS..258...11W}. The catalog offers comprehensive photometric coverage spanning the 2 square degree COSMOS field. Its extensive and continuous coverage across 40 bands, ranging from the u band to IRAC ch4, is instrumental in constructing an overdensity map. To assess the consistency of photometric redshifts with our narrow-band redshift range, we examine the cross match 1331 emitters with the LePhare photo-z from the catalog. By comparing each source's photo-\(z\) within the 68\% confidence interval (\(lp\_zPDF\_l68\) to \(lp\_zPDF\_u68\)) to the targeted narrow-band redshift range, we found that 799 sources fit within this narrow-band redshift range.

Following \citet{2013A&A...556A..55I}, galaxies are classified as quiescent if they satisfy the rest-frame color criteria in the NUV\textendash r vs.\ r\textendash J diagram, defined as \( M_{\text{NUV}} - M_r > 3(M_r - M_J) + 1 \) and \( M_{\text{NUV}} - M_r > 3.1 \). For our analysis, we consider only those galaxies within a redshift slice \( \left| \Delta z \right| < 2\sigma_z (1 + z) \), where \( z=1.471 \) and \( \sigma_z = 0.015 \). We limited the redshift range 1.43 to 1.51 which is consistent with narrow band redshift range. Within this defined sample, we identified 360 quiescent galaxies that possess stellar masses greater than or equal to $\mathrm{10^{9} M_{\odot}}$.

\subsection{JWST NIRCam Imaging}\label{sec:nircam}
We utilize high-resolution JWST Near Infrared Camera (NIRCam) imaging data from the COSMOS-Web survey \citep{2023ApJ...954...31C} taken during two epochs (Epoch 1: January 5/6, 2023; Epoch 2: April/May 2023) shown by green line in Figure \ref{fig:cosmic_structure}. The NIRCam on the JWST provides crucial data for this study, spanning a wavelength range of 1.0 to 5.0 microns. 

Employing four NIRCam filters (F115W, F150W, F277W, and F444W), the survey achieves a 5-sigma point-source sensitivity between 27.5 and 28.2 magnitudes over a contiguous 0.54 square degree area.

For our specific analysis, we leveraged the available COSMOS-Web NIRCam imaging that covers our area of interest—the filamentary overdensity of [OII]-selected galaxies at z $\sim$ 1.5. We reduced the raw data taken from MAST, by using the JWST data reduction pipeline (v1.10.0) with several customized steps as presented in \citet{morishita23}. The images used in this work are processed with CRSD context version \texttt{jwst\_1080.pmap}. The final mosaic images are aligned to the WCS of the aforementioned COSMOS catalog and set to the pixel size of 0.0315\,arcseconds per pixel.

\section{\textbf{Methods}} \label{sec:method}

\subsection{Star Formation Rate and Stellar Mass} \label{subsec:sfr_mass_analysis}

\subsubsection{Star Formation Rate}

We determine the emission line luminosity by calculating the line flux from the flux densities observed in both narrowband (NB) and broadband (BB) filters \citep{2020PASJ...72...86H}. 

We estimate the SFRs of [OII] emitters by using the \citet{1998ARA&A..36..189K}
calibration assuming solar abundances and the Salpeter \citep{1955ApJ...121..161S} IMF :
\begin{align}
           & {\rm SFR}_{\rm [OII]}{\rm [M_{\odot} yr^{-1}]}=(1.4 \pm 0.4) \times 10^{-41}\cdot \frac{L_{\rm [OII]}}{\rm erg\ s^{-1}}
\end{align}
For consistency, we convert the \citet{1955ApJ...121..161S} IMF to \citet{Chabrier_2003} by dividing the previous calibration by a factor of 1.7 \citep{Pozzetti_2007}.

We correct dust attenuation to the measured [OII] luminosity by adopting SED-based dust attenuation $A_v$ (see Sec.~\ref{sed:sed}) with the extinction curve of \citet{2000ApJ...533..682C}.

\subsubsection{SED Fitting with CIGALE}\label{sed:sed}
We leverage the wealth of photometric data from the COSMOS 2020 catalog and employ the Code Investigating GALaxy Emission (CIGALE) for spectral energy distribution (SED) fitting. Using CIGALE \footnote{\url{https://cigale.lam.fr/}} \citep{2005MNRAS.360.1413B, 2009A&A...507.1793N, 2019A&A...622A.103B}, we fit the multi-wavelength data of our emission line galaxies to determine stellar mass (M$_*$) and dust extinction ($A_v$). We assume a delayed star-formation history for the galaxies and generate SED templates that incorporate stellar emission based on stellar population models \citep{2003MNRAS.344.1000B}. Our analysis adopts a modified starburst attenuation law \citep{2000ApJ...533..682C} that accounts for differential reddening of different age stellar populations. We use a range for the color excess of the nebular lines, \( E(B - V )_{\rm line} \), from 0 to 1 in steps of 0.06, and set the ratio between stellar and nebular attenuation (\textit{f}) to 0.44. The uncertainty surrounding the \textit{f}-factor at high redshift remains considerable, as highlighted by various studies \citep{2013ApJ...777L...8K, 2014ApJ...788...86P, 2019PASJ...71....8K,2020ApJ...902..123R}. Moreover, modifying the \textit{f}-factor to 0.7 does not alter the primary conclusions of our analysis. The SEDs assume an initial mass function (IMF) from \citet{Chabrier_2003} and consider stellar metallicities of 0.004, 0.008, and 0.02. The $e$-folding times of the main stellar population range from 1 to 30 Gyrs, while the ages of the primary stellar population are determined based on the redshift of the galaxy.

\subsection{Morphological Analysis} \label{subsec:morph_analysis}

To investigate the structures of our [OII] emitters, we exploit the high-resolution imaging provided by the JWST NIRCam F150W filter when available. This filter, centered at $\sim1.5$\,microns, captures the rest-frame optical light of galaxies at \( z \approx 1.5 \), making it ideal for detailed morphological studies of our [OII] samples. We examine the morphological characteristics of our galaxy sample using {\tt Statmorph}, a Python tool \citep{2019MNRAS.483.4140R}.  {\tt Statmorph} calculates effective radius ($R_{\rm eff}$), and several morphological indices, including concentration, asymmetry, smoothness \citep{2008ApJ...672..177L, 2014ARA&A..52..291C}, and Gini-M20 coefficients \citep{2008ApJ...672..177L, 2015MNRAS.451.4290S,2015MNRAS.454.1886S}. These indices are crucial for identifying key galaxy features, including clumpiness, interaction traces, and merger indications.

For our analysis, we prepared 100 × 100 pixel sections of the galaxy images for input into {\tt Statmorph}, accompanied by the corresponding weight images and segmentation maps to isolate the target galaxies. We also provided the NIRCAM F150W Point Spread Function (PSF), generated by WebbPSF \citep{2014SPIE.9143E..3XP}, although it is important to note that PSF is not used in nonparametric morphological measurements. Some measurements are marked as unreliable by {\tt Statmorph} because of factors such as insufficient skyboxes, low S/N, or unsuccessful Sérsic fitting. We do not include these marked objects in the results.

As a sanity check of the measured sizes, we employed {\tt GALFIT} \citep{2002AJ....124..266P}, a widely-used fitting tool that relies on S\'ersic light profiles. A comparison between the size measurements from {\tt Statmorph} and {\tt GALFIT} for the unflagged galaxies in our sample show a strong agreement, confirming the reliability of these two independent methods in our morphological study.

\section{\textbf{Large-scale structures traced by [OII] Emitters}} \label{sec:overdensity}

The spatial distribution of the [OII] emitters is shown in Figure~\ref{fig:cosmic_structure} (top). We observe a clear overdensity of [OII] emitters in a filament-like structure that extends from the north to southwest. Additionally, there is an extension of this filamentary structure to the east. To the northwest and southwest of the overdense core center, there are two regions of relatively high density. These features seem to be connected to the core of the structure, comprising a large-scale structure spanning over 50 cMpc, indicative of a cosmic web structure at this redshift.

The discovered overdensity highlights the clustering of star-forming galaxies, possibly indicating the initial phase of galaxy cluster formation. This contrasts with the nearby universe, where dense environments have inactive red galaxies and star-forming galaxies are in low-density environments. We calculate the local density environment using the Nth nearest neighbor approach. For defining local density, we consider a radius encompassing the 10th nearest neighbor object ($\Sigma_{\mathrm{10th}}$), combining both [OII] emitters and quiescent galaxies. This density distribution pattern is found consistent even when varying the Nth value in our analysis in the range of 5-10. 

Based on the measured local density, we here classify galaxies that are within the  $r<1.\!'5$ from the densest central coordinates $(\alpha = 9^{\text{h}}59^{\text{m}}54^{\text{s}}, \delta = +02^{\circ}20'00.75'')$ as core members. 
Those situated between $1.\!'5$ and 5$'$ are classified as outskirt galaxies. Figure~\ref{fig:cosmic_structure} (bottom-left) shows the zooming view of the core and outskirts region.  Those in the lower 45th percentile of the $\Sigma_{\mathrm{10th}}$ distribution are classified as field galaxies.
The classification of field galaxies remains robust even when considering the lower 25th percentile. 

We evaluate the overdensity of [OII] emitters in the core. 
The observed average surface density of 1447 [OII] emitters in the total survey area of 4932 arcmin$^{2}$, is $(2.93 \pm 0.08) \times 10^{-1} \, \text{arcmin}^{-2}$ where the uncertainty reflects the Poisson noise. In the most overdense core region, spanning roughly 7 arcmin$^{2}$ in which 22 [OII] emitters are enclosed, the number density is $(3.11 \pm 0.66), \text{arcmin}^{-2}$, signifying an 11-fold increase compared to the overall surveyed region. The probability of the observed overdensity resulting from Poisson fluctuations is  $\sim 10^{-15}$. This suggests a highly unlikely scenario under the assumption of a Poisson distribution, indicating that overdensity is likely due to a significant underlying structure. 

From the Chandra COSMOS Legacy Survey \citep{2016ApJ...819...62C}, which catalogs 4016 X-ray point sources across 1.5 deg$^2$ of the COSMOS field, we identified 29 spectroscopically confirmed AGN between redshifts 1.45 and 1.49, and associated host galaxies as reported by \cite{2016ApJ...817...34M}. The bottom-right panel of Figure~\ref{fig:cosmic_structure}  presents the local density $\Sigma_{\mathrm{10th}}$ [1/Mpc$^{2}$] for different galaxy populations, including the X-ray AGNs. The median value of the X-ray AGN $\Sigma_{\mathrm{10th}}$ distribution positions itself within the interquartile range of 50--75\%. Of the 29 X-ray AGNs, 11 are situated within the highest density quartile (75--100\%), 7 fall within the 50--75\% range, 2 are categorized between 25--50\%, and 9 are found within the least dense quartile (0--25\%). Within these distributions, four Xray AGNs reside in the outskirts, while the core contains only one. Additionally, cross-matching within a 1 arcsecond radius identified two of our [OII] emitters as X-ray AGNs, with one belonging to the outskirts sample. These have been excluded from our [OII] emitter sample for subsequent analysis. It is generally acknowledged that AGNs can significantly influence star formation in galaxies \citep{2014MNRAS.445..581H,2015ARA&A..53...51S, 2017MNRAS.472..949B, 2018MNRAS.476..979P}. While existing observations at low redshifts indicate a lower prevalence of AGN activity in dense environments \citep{1978MNRAS.183..633G, 2010MNRAS.404.1231V, 2018ApJ...869..131M}, suggesting a potential reversal of this relationship at higher redshifts \citep{2009ApJ...691..687L, 2010MNRAS.407..846D, 2016ApJ...825...72A, 2017MNRAS.470.2170K, 2022ApJ...941..134H}. A comprehensive examination of the link between AGN activity and environmental density is beyond the scope of this study.

\begin{deluxetable*}{ccccccc}
\tablecaption{ Morphological Parameters \label{tab:data_used}}
\tablewidth{0pt}
\tabletypesize{}
\tablehead{
\colhead{} & \colhead{N$_{\mathrm{[OII]}}^{\textcolor{blue}{a}}$} & \colhead{N$_{\mathrm{JWST}}^{\textcolor{blue}{b}}$} & \colhead{N$_{\mathrm{flag}}^{\textcolor{blue}{c}}$} & \colhead{Disturbance Fraction (\%)}& \colhead{$\mathrm{R_{eff}}$ (kpc)} & \colhead{Sersic Index} 
}

\startdata
Whole & 1447 & 368 & 10 &  -- & -- & -- \\
\tableline 
Core & 22 & 22 & 0 & 50\% $\pm$ 9\% & $2.582^{+0.326}_{-0.462}$ & $0.865^{+0.214}_{-0.123}$ \\
\tableline 
Outskirts & 59 & 40 & 1  &36\% $\pm$ 8\%  & $2.358^{+0.332}_{-0.174}$ & $0.868^{+0.233}_{-0.173}$ \\
Mass Control & 46 & 32 & 0 &41\% $\pm$ 9\% & $2.439^{+0.272}_{-0.226}$ & $0.854^{+0.115}_{-0.302}$ \\
\tableline 
Field & 623 & 103 & 5 &27\% $\pm$ 4\% & $2.019^{+0.157}_{-0.211}$  & $1.029^{+0.073}_{-0.087}$ \\
Mass Control & 539 & 89 & 5 &21\% $\pm$ 5\%  & $1.875^{+0.243}_{-0.095}$  & $0.988^{+0.092}_{-0.059}$ \\
\tableline 
\enddata
\tablecomments{\\
$^a$Number of [OII] emitters.\\
$^b$Number of galaxies observed in the available COSMOS Web footprint\\
$^c$Number of galaxies flagged by {\tt Statmorph}.
}
\end{deluxetable*}

\section{\textbf{Results}} \label{sec:results}

\subsection{Morphological Disturbances in Different Environments}

We aim to investigate the morphological properties of galaxies residing in different environments: overdense core, outskirts, and field, as described in the previous section. Utilizing the COSMOS Web footprint from the initial two epochs of the COSMOS-Web survey, we identify 368 with NIRCam data coverage out of 1447 total galaxies.  This includes all 22 galaxies identified in the core areas, 40 out of 59 galaxies in the outskirts, and 103 out of 623 field galaxies. Table \ref{tab:data_used} provides a brief summary of the data and morphological parameters.

\begin{figure*}[ht]
  \centering
  \includegraphics[width=0.99\textwidth]{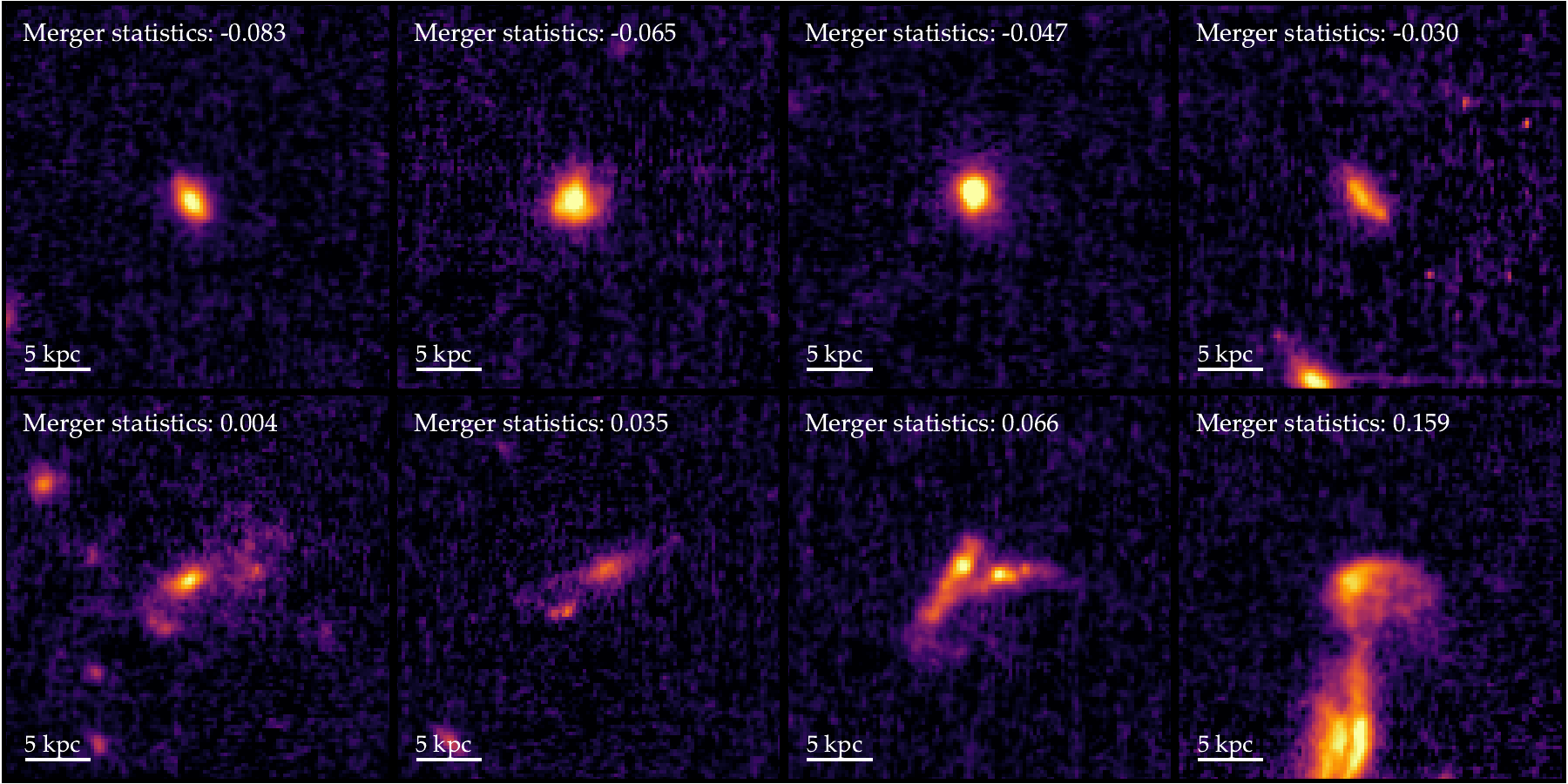}
  \caption{Example galaxy images from JWST NIRCam observations in the F150W band, in the order of increasing merger statistic values. Each panel is annotated with its corresponding merger statistic value, quantifying the degree of morphological disturbance or indications of merging activity. 
  These images exhibit a spectrum of morphological disruptions in galaxies, ranging from minor to more significant.}
  \label{fig:morphology}
\end{figure*}

In Figure \ref{fig:morphology}, we show a subset of galaxies imaged by JWST NIRCam in the F150W band, which corresponds to the rest-frame optical. 
To quantify the observed morphological differences across different environments, we characterized the apparent morphologies of the galaxy samples using {\tt Statmorph}. {\tt Statmorph} provides a flag when the morphology parameters cannot be accurately determined. We identified a fraction of our sample for which {\tt Statmorph} failed to measure non-parametric morphology statistics. Specifically, of the 368 galaxies, 358 yielded reliable non-parametric measurements. This includes all 22 galaxies in the core regions, 39 out of 40 in the outskirts, and 98 out of 103 in the field regions (Table~\ref{tab:data_used}). Galaxies flagged by {\tt Statmorph}, indicating the presence of artifacts or inaccurate Gini segmentation map estimations, were excluded from the analysis.

Visual inspection of the galaxy samples revealed variations in their morphological structures across different environmental contexts. Notably, galaxies residing in the overdense core exhibit the most substantial morphological disturbances, with outskirts galaxies showing moderate disturbances in comparison, and field galaxies displaying the least disturbance. We further explored these morphological variations using the Gini-M20 statistic, which is an effective indicator for identifying potentially disturbed galaxies.

Our study utilized the Gini-M20 statistic, a comprehensive metric sensitive to various phenomena such as minor mergers, gas-poor mergers, and low signal-to-noise ratios \citep{2004AJ....128..163L, 2008ApJ...672..177L,2010MNRAS.404..575L, 2016MNRAS.458..963P}. 
In our analysis, we utilize the Gini-M$\mathrm{_{20}}$ `merger statistic', $\mathrm{S(G, M_{20}) = 0.139M_{20} + 0.990G - 0.327}$ \citep{2015MNRAS.451.4290S}, which we derived from the {\tt Statmorph} output \textit{gini\_m20\_merger} \citep{2019MNRAS.483.4140R}. 
We consider galaxies with \textit{gini\_m20\_merger} $\geq$ 0 as disturbed. Our findings reveal that 50\% $\pm$ 9\% of core galaxies exhibit signs of disturbance, in contrast to 36\% $\pm$ 7\% in outskirts and 27\% $\pm$ 4\% in the field population. These uncertainties were determined using bootstrap resampling, with the 16th and 84th percentiles. 

\begin{figure*}[htp]
\centering
\includegraphics[width=0.5\linewidth]{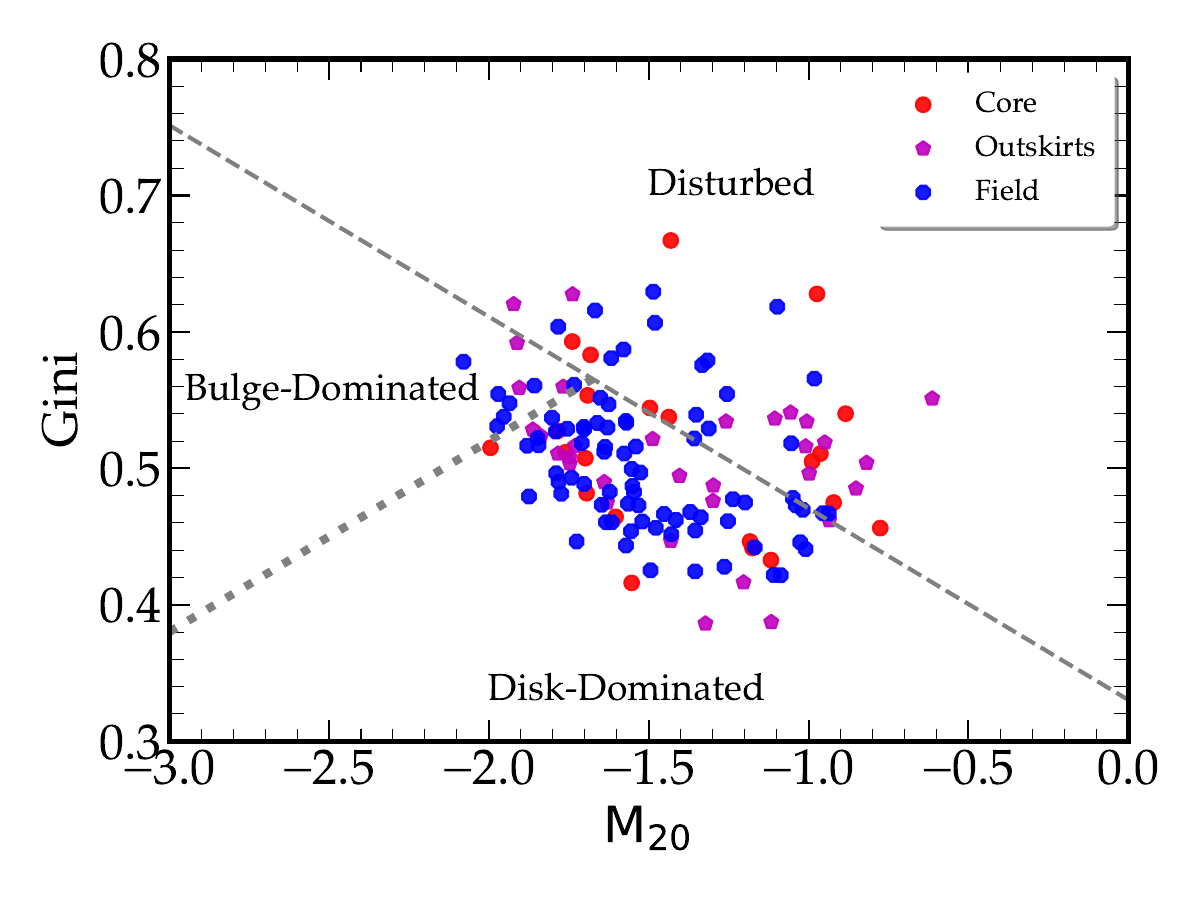} %
\vspace{4pt} 
\includegraphics[width=1.0\linewidth]{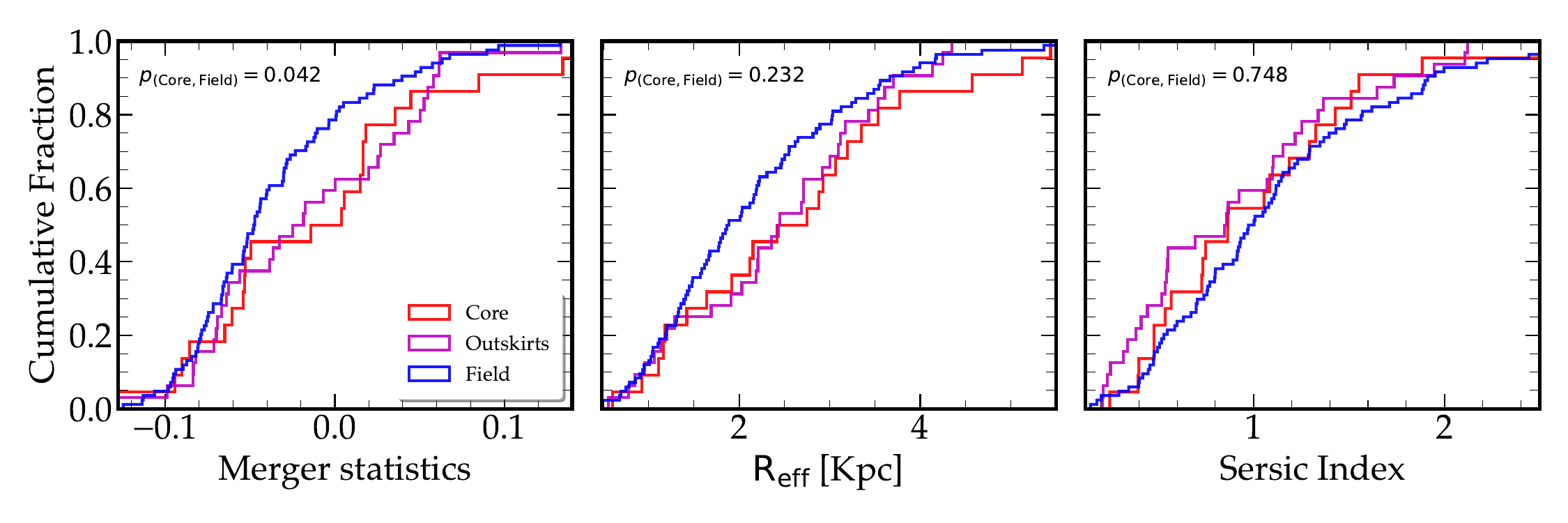} %
\caption{Top: Displays the distribution of core, outskirts, and field galaxies within the Gini--M$_{20}$ space, represented by red, magenta, and blue colors, respectively. The gray dotted line indicates the transition between disk-dominated regions and bulge-dominated regions. The dashed line identifies the separation line for merger statistics, situating potential merger candidates above this line, as discussed in \citet{2015MNRAS.451.4290S} and \citet{2019MNRAS.483.4140R}. Bottom: The bottom panels show the cumulative distribution functions (CDFs) for core (red), outskirts (magenta), and field (blue) galaxies across various morphological parameters: merger statistics, effective radius (R$_{\text{eff}}$ in kpc), and Sersic index. These CDFs provide a statistical representation of the environmental dependence of galaxy structure across different environments.}
\label{fig:morphology_quantity}
\end{figure*}

\subsubsection{Quantitative comparison of morphologies in different environments}

Distinguishing the intrinsic effects of stellar mass from the specific effects of the environment on galaxy morphology requires mitigation of potential biases. We address this challenge by constructing control samples.
Specifically, we implement this by restricting our analysis to galaxies within a similar mass range of the core, outskirts and field samples. This approach ensures that the observed variations primarily reflect environmental influences rather than mass-related differences. Additionally, we ensure that our samples are mass-complete based on the criteria established for the COSMOS2020 catalog by \citet{2022ApJS..258...11W}. This mass completeness is calculated using LePhare based on IRAC channel 1 magnitude limits, following the method described by \citet{2010A&A...523A..13P}. The mass completeness limit, applicable to both star-forming and quiescent galaxies, is presented in Figure 20 of \citet{2022ApJS..258...11W} and corresponds to ($\mathrm{4 \times 10^{8} M_{\odot}}$) for star-forming galaxies at $z\sim$ 1.5. We only consider galaxies that have stellar masses within the mass completeness limit ($\mathrm{4 \times 10^{8} M_{\odot}}$) and upper mass limit the core sample ($\mathrm{1.4 \times 10^{10} M_{\odot}}$). This overdensity is dominated by lower mass galaxies,  similar to the USS1558 protocluster at $z=2.53$ \citep{2018MNRAS.473.1977S}, which is also dominated by lower mass galaxies. The limiting dust-corrected SFR of the sample is $1.4\ \mathrm{M_\odot  yr^{-1}}$.
The number of control samples across the various populations is presented in Table \ref{tab:data_used}.

The distribution of galaxy morphologies within the Gini–M20 classification space is illustrated in Figure~\ref{fig:morphology_quantity} (top panel), where demarcation lines distinguish between bulge-dominated and merger-influenced galaxies. We found that 50\% $\pm$ 9\% of core galaxies, 41\% $\pm$ 9\% of outskirts and 21\% $\pm$ 5\% of field exhibit signs of disturbance, as indicated by merger statistics with \( S(G, M_{20}) \geq 0 \).

The cumulative distribution function for the merger statistic is plotted in the bottom left panel of Figure~\ref{fig:morphology_quantity}. A Kolmogorov-Smirnov (KS) test between core and field populations yields a \textit{p}-value of 0.042, suggesting a statistically significant difference in the frequency of disturbed morphologies between these environments. Outskirts galaxies also present a median merger statistic higher than that of the field, with a \textit{p}-value of 0.138, hinting at a possible environmental influence on galaxy interactions, although this observation does not reach a high level of statistical significance.  
This aligns with the prevailing understanding that denser environments promote conditions favorable for galaxy interactions and mergers. This observation is in line with previous studies \citep[e.g.,][]{2008ApJ...672..177L, 2012ApJ...750...93P} that reported heightened merger activities in denser areas. The core regions, which are parts of dynamically active regions, are more susceptible to such interactions, leading to observed morphological disturbances. The outskirts, representing a transitional zone, show an intermediate level of disturbance. This could suggest the influence of ongoing environmental processes, such as accretion and the effects of the cluster's gravitational potential, on galaxy morphology. Field galaxies, residing in less dense environments, exhibit the least morphological disturbances, underscoring the reduced impact of dense environments on their evolution. This is consistent with the expectations for galaxies in lower density regions, where interactions and mergers are less frequent.

In the bottom-middle panel of Figure~\ref{fig:morphology_quantity}, the cumulative distribution function for the effective radius is shown for the core (red), outskirts (magenta), and field (blue) galaxies. The median effective radius of core galaxies is measured at \(2.58^{+0.33}_{-0.46}\) kpc, which is a 37\% increase, corresponding to a \(1.65\sigma\) deviation, compared to the field galaxies, which have a median effective radius of \(1.88^{+0.24}_{-0.10}\) kpc. Although this indicates a tendency for larger sizes in core galaxies, the difference does not attain a high level of statistical significance, as evidenced by a KS test \textit{p}-value of 0.234. Outskirts galaxies exhibit a median effective radius of \(2.44^{+0.27}_{-0.23}\) kpc, which is
approximately 30\% larger than that of field galaxies.

The bottom-right panel of Figure~\ref{fig:morphology_quantity} presents the cumulative distribution function for the Sérsic index. Core galaxies exhibit a median Sérsic index of \(0.87^{+0.21}_{-0.12}\), with outskirts galaxies displaying a similar median of \(0.85^{+0.12}_{-0.30}\). In comparison, the field sample demonstrates a slightly elevated median Sérsic index of \(0.99^{+0.09}_{-0.06}\). The KS test yields a \textit{p}-value of 0.748 when comparing the core and field galaxies, indicating that the differences in the Sérsic index among core, outskirts, and field galaxies are not statistically significant within a \(1\sigma\) confidence level. This suggests that variations in galaxy morphologies across these environments are not substantial.

\subsection{Effect of the Environment on Star-forming Activity}

\begin{figure*}[htp]
\centering
\includegraphics[width=0.8\linewidth]{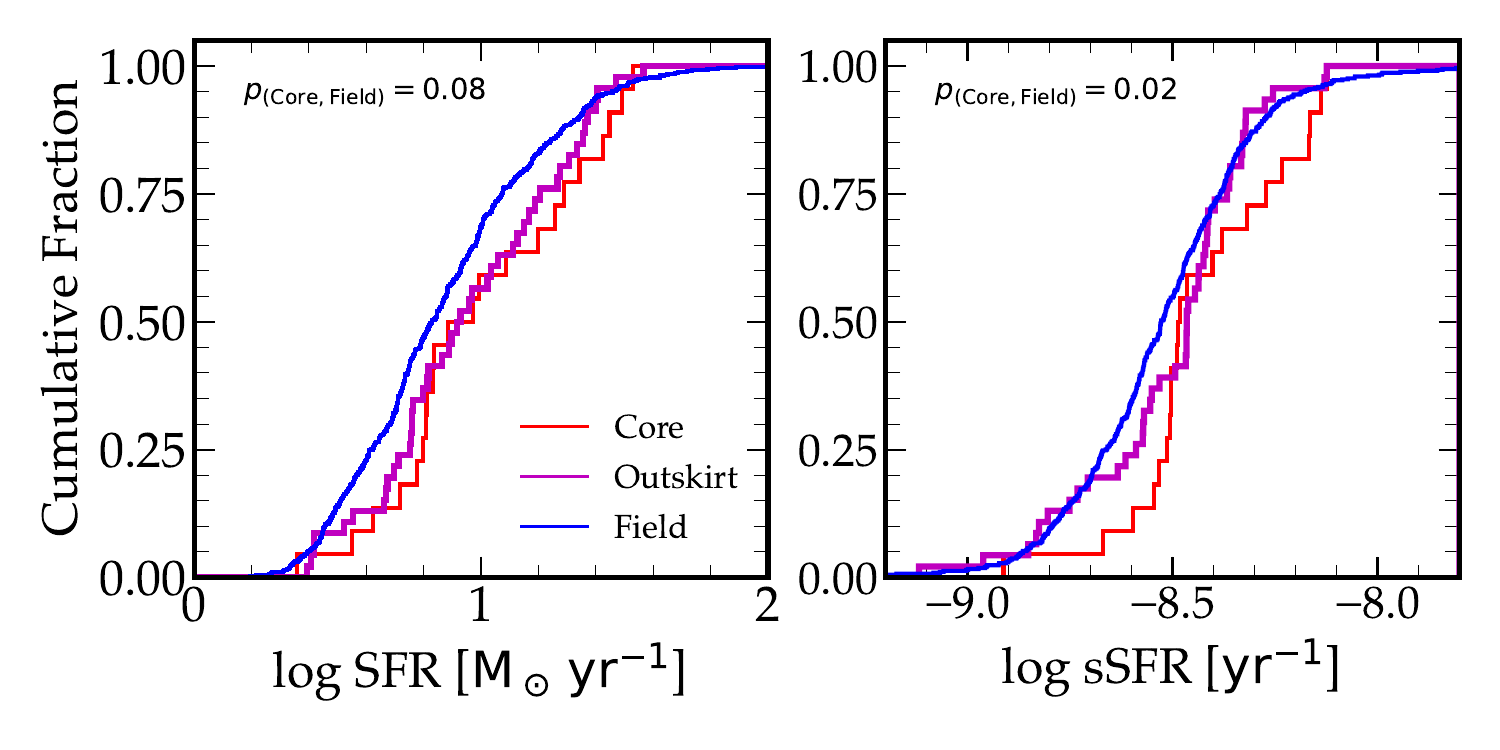} 
\vspace{5pt}
\includegraphics[width=1\linewidth]{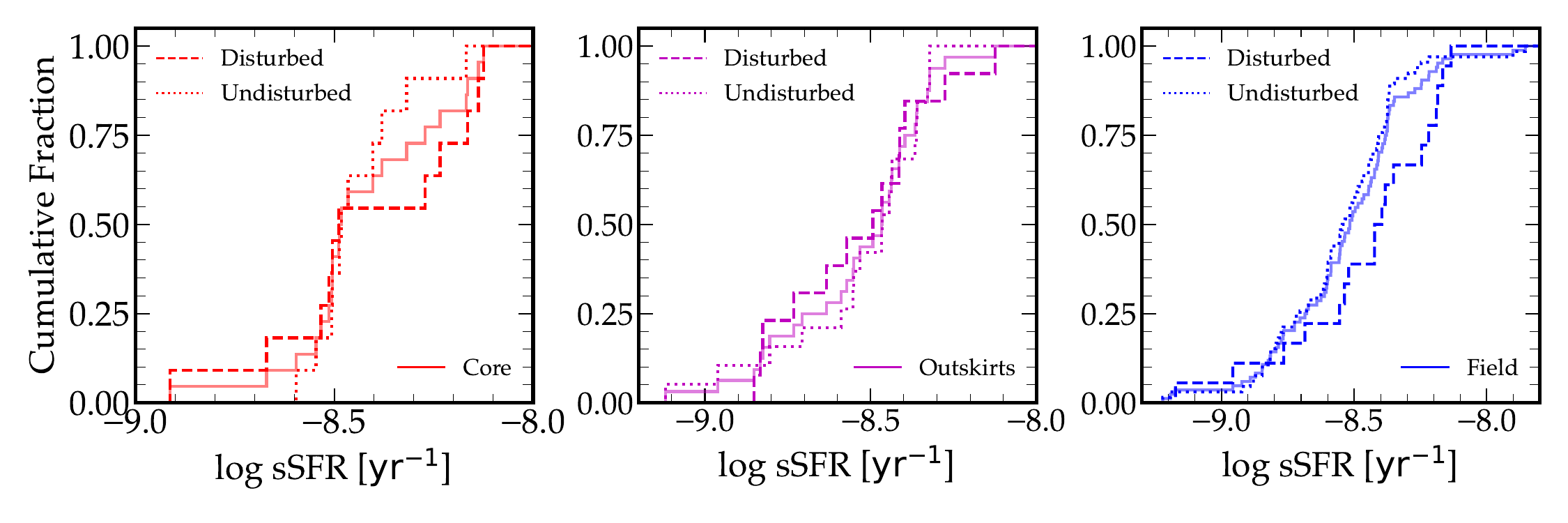} 
\caption{Top: Cumulative distribution functions (CDFs) of star formation rate (SFR) and specific star formation rate (sSFR) for galaxies within the core, outskirts, and field control samples. Bottom: CDF of sSFR for disturbed ($\mathrm{S(G, M_{20}) \ge 0}$), undisturbed ($\mathrm{S(G, M_{20}) < 0}$) galaxies based on Gini-M$_{20}$ statistic across different environments.}
\label{fig:cdf_plot}
\end{figure*}

We investigate the star formation activity across the core, outskirts, and field mass control samples. Our samples here are not limited to the COSMOS Web footprint. 
We compared the star formation rate (SFR) and specific star formation rate (sSFR) for each population. Figure \ref{fig:cdf_plot} (top) presents the cumulative distribution of SFR and sSFR.

The median SFR, measured using the [OII] luminosity, reveals a trend across different environments. Core galaxies exhibit the highest median SFR of $\mathrm{SFR_{[OII]}} = 8.52^{+3.70}_{-1.68}\ \mathrm{M_\odot yr^{-1}}$, indicative of a significant level of ongoing star formation, closely followed by the outskirts galaxies with $8.36^{+1.73}_{-1.07}\ \mathrm{M_\odot yr^{-1}}$. Field galaxies display the lowest median SFR of $6.75^{+0.30}_{-0.33}\ \mathrm{M_\odot yr^{-1}}$, reflecting reduced star-forming activity. This pattern suggests that denser environments may be more conducive to star formation. KS test yields a \textit{p}-value of 0.08 when comparing the core to the field and 0.12 for the outskirts to the field, indicating that the differences in SFR between core and field galaxies are noticeable but not statistically significant at the standard threshold of 0.05. While the overall results remain consistent, the significance level increases, with \textit{p}-values falling below 0.05, when focusing on field galaxies below the 35th percentile.

Given the observed variations in SFR across different environments, we next examined the specific star formation rates (sSFR) to further understand the influence of environmental density on star formation efficiency.  The core's median sSFR is $3.31^{+0.16}_{-0.08} \times 10^{-9} \, \mathrm{yr}^{-1}$, while the outskirts have a median of $3.39^{+0.24}_{-0.00} \times 10^{-9} \, \mathrm{yr}^{-1}$, and the field has $2.96^{+0.07}_{-0.00} \times 10^{-9} \, \mathrm{yr}^{-1}$. The KS test indicates a more significant difference in sSFR between the core and field with \textit{p}-value of 0.02. Outskirts and field sample have \textit{p}-value of 0.05, while the core and outskirts do not differ significantly (\textit{p}-value of 0.33). 

In Figure \ref{fig:cdf_plot} (bottom), we examine the specific star formation rates (sSFR) of galaxies, distinguishing between disturbed (dashed lines) and undisturbed (dotted lines) galaxies within core (red), outskirts (magenta), and field (blue) environments. Across all environments, the disturbed galaxies exhibit a broader and slightly skewed distribution towards higher sSFR values, while the undisturbed galaxies display a narrower distribution. This pattern may indicate that within dense environments, disturbed galaxies are likely to experience heightened star formation activities compared to their undisturbed counterparts. However, the differences in sSFR between disturbed and undisturbed galaxies do not show statistical significance across the core, outskirts, and field samples. This observation indicates that while morphological disturbances may correlate with variations in star formation activity, these variations are not distinct enough to reach statistical significance within the scope of our study.

\section{\textbf{Discussion and Summary}} \label{sec:discussion}

This study has presented a pilot analysis of [OII] emitters within the COSMOS UltraDeep field at \( z \sim 1.5 \), leveraging narrowband data from the Hyper Suprime-Cam Subaru Strategic Program and high-resolution imaging from the JWST NIRCam as part of the COSMOS-Web survey. Our investigation reveals large filamentary structures traced by [OII] emitters, surrounding an extremely overdense core. 
By taking advantage of the two unique datasets, we have shed light on the critical role of environment on galaxy evolution at $z\sim $1.5.

Our findings suggest that galaxies in denser environments, particularly core regions, not only demonstrate higher star formation rates but also exhibit more structural disturbances, implying active galaxy formation processes such as gas cooling and mergers. The observed trends hint at a possible link between the level of structural disturbance and the extent of enhanced star formation \citep[e.g.,][]{2016MNRAS.455..295H}.  
The enhanced star-forming activities within this overdense filamentary structure could be due to the increased efficiency of gas accretion. Such a process may result in a higher supply of cold gas to galaxies, subsequently elevating their star formation rates. Furthermore, some simulations also suggest that the influx of cold gas along the cosmic web's filaments supplies the necessary fuel for galaxies, thereby driving up their star formation rates at high redshifts \citep[e.g.,][]{2005MNRAS.363....2K, 2006MNRAS.368....2D, 2008ApJ...688..789G}.
Continuous gas inflow is likely to contribute to disk instability \citep{2009Natur.457..451D}, indicating environmental variations in gas accretion mechanisms. Such mechanisms may result in a higher prevalence of disturbed or irregular galaxy morphologies in dense areas than in field regions, which is consistent with our findings. Unlike galaxy clusters in the local universe that are dominated by their passive population, this structure appears to be in an early  stage of vigorous assembly with a higher gas fraction, as reported in  \citep[e.g.,][]{2020A&A...641L...6D, 2020A&A...635L..10C}.
Therefore, interaction with the hot gas may not yet be significant in this structure, which could potentially not affect star-forming galaxies through mechanisms such as ram pressure stripping. 
Several studies indicate that star formation in dense environments at high redshifts is enhanced compared to field galaxies, attributed to the increased rates of gas accretion and mergers \citep{2014MNRAS.437..437A,2021ApJ...919...51M,2022A&A...662A..33L,2024ApJ...961...39S,2024MNRAS.52710221P}.
\citet{2018MNRAS.473.1977S} also found that star-forming galaxies in dense groups at $z\sim$ 2.5 had elevated specific star formation rates (sSFRs) relative to their counterparts in less dense environments, likely due to their substantial HI gas reserves in the proto-cluster \citep{2016ApJ...817..160L, 2017ApJ...839..131C,2017MNRAS.467.3951M}. Our overdense structure may also have a large amount of HI gas. Future observations of molecular gases are essential to confirm the HI gas content in this structure and further understand its influence on galaxy evolution.

Our study has focused on analyzing the morphology and star formation activities of star-forming galaxies in different environments. While the exact impact of our results is yet to be fully understood, we found evidence of environmental impact on the morphology and star-formation. Our key findings can be summarized as follows:

\begin{enumerate}
    \item The [OII] emitters have unveiled a substantial filamentary structure over 50 cMpc, surrounding a core of extreme galaxy overdensity, with a galaxy number density approximately $11\times$ higher than the field average.
    The core of this structure, characterized by a pronounced overdensity of star-forming galaxies, potentially marks an early phase of galaxy cluster formation. This contrasts with typical associations in the nearby universe, where dense environments harbor inactive red galaxies, and star-forming galaxies are more prevalent in less dense areas.
    
    \item A significant increase in morphological disturbances has been observed in core region galaxies compared with those in outskirts and field environments. This finding highlights the profound impact of environmental density on galaxy morphology, especially in denser areas where interactions and mergers become more frequent.
    
    \item Core galaxies exhibited higher SFR and sSFR compared to field galaxies. This, along with similar stellar mass distributions across environments, emphasizes the role of environmental density in influencing star formation efficiency, suggesting accelerated processes such as gas cooling and mergers in denser environments.
    
    \item While no statistically significant correlation was found between structural disturbance and sSFR, a trend was observed where galaxies with greater degrees of disturbance displayed a marginally higher sSFR. This trend, observed across the core, outskirts, and field regions, suggests a potential connection between structural disturbance and enhanced star formation.
    
\end{enumerate}

Those galaxies in the overdense core may be undergoing merger and are thought to play a major role in the growth of massive galaxies. It is important to recognize that the findings of this study regarding the impact of environment on galaxies within this overdense structure might not be universally applicable to every cluster or overdense region at this epoch in the universe. Outcomes could significantly vary based on the developmental stage of each cluster, even if they share the same redshift. Clearly, more extensive samples of galaxies at this and other redshifts are needed to generalize these results and to validate the observed differences in morphology. In future work, we aim to broaden the scope of our research by incorporating a larger dataset, including the additional data from ongoing and upcoming surveys. This expansion will not only enhance our sample size but also enable more robust statistical inferences. This will allow for a more comprehensive analysis that considers additional variables influencing galaxy evolution, such as interactions between galaxies and the effects of AGN feedback on star formation and morphology. This approach will enrich our understanding of the complex interplay between environmental conditions and galaxy evolution.

\section*{Acknowledgements}
We are grateful to the anonymous reviewer for their insightful comments and suggestions that improved the manuscript. This work is based on data collected at the Subaru Telescope and retrieved from the HSC data archive system, which is operated by Subaru Telescope and Astronomy Data Center (ADC) at NAOJ.

R.L. was previously supported by MEXT and currently by Graduate Program on Physics for the Universe (GPPU), Tohoku University. R.L. acknowledges support from JST SPRING, grant No. JPMJFS2102, and KAKENHI International Leading Research (Grant Number 22K21349). T.K. acknowledges financial support from JSPS KAKENHI Grant Number 18H03717.


\vspace{5mm}

\bibliography{sample631}{}

\begin{thebibliography}{}
\expandafter\ifx\csname natexlab\endcsname\relax\def\natexlab#1{#1}\fi
\providecommand{\url}[1]{\href{#1}{#1}}
\providecommand{\dodoi}[1]{doi:~\href{http://doi.org/#1}{\nolinkurl{#1}}}
\providecommand{\doeprint}[1]{\href{http://ascl.net/#1}{\nolinkurl{http://ascl.net/#1}}}
\providecommand{\doarXiv}[1]{\href{https://arxiv.org/abs/#1}{\nolinkurl{https://arxiv.org/abs/#1}}}

\bibitem[{{Aihara} {et~al.}(2018){Aihara}, {Arimoto}, {Armstrong}, {Arnouts}, {Bahcall}, {Bickerton}, {Bosch}, {Bundy}, {Capak}, {Chan}, {Chiba}, {Coupon}, {Egami}, {Enoki}, {Finet}, {Fujimori}, {Fujimoto}, {Furusawa}, {Furusawa}, {Goto}, {Goulding}, {Greco}, {Greene}, {Gunn}, {Hamana}, {Harikane}, {Hashimoto}, {Hattori}, {Hayashi}, {Hayashi}, {He{\l}miniak}, {Higuchi}, {Hikage}, {Ho}, {Hsieh}, {Huang}, {Huang}, {Ikeda}, {Imanishi}, {Inoue}, {Iwasawa}, {Iwata}, {Jaelani}, {Jian}, {Kamata}, {Karoji}, {Kashikawa}, {Katayama}, {Kawanomoto}, {Kayo}, {Koda}, {Koike}, {Kojima}, {Komiyama}, {Konno}, {Koshida}, {Koyama}, {Kusakabe}, {Leauthaud}, {Lee}, {Lin}, {Lin}, {Lupton}, {Mandelbaum}, {Matsuoka}, {Medezinski}, {Mineo}, {Miyama}, {Miyatake}, {Miyazaki}, {Momose}, {More}, {More}, {Moritani}, {Moriya}, {Morokuma}, {Mukae}, {Murata}, {Murayama}, {Nagao}, {Nakata}, {Niida}, {Niikura}, {Nishizawa}, {Obuchi}, {Oguri}, {Oishi}, {Okabe}, {Okamoto}, {Okura}, {Ono}, {Onodera}, {Onoue}, {Osato}, {Ouchi}, {Price}, {Pyo},
  {Sako}, {Sawicki}, {Shibuya}, {Shimasaku}, {Shimono}, {Shirasaki}, {Silverman}, {Simet}, {Speagle}, {Spergel}, {Strauss}, {Sugahara}, {Sugiyama}, {Suto}, {Suyu}, {Suzuki}, {Tait}, {Takada}, {Takata}, {Tamura}, {Tanaka}, {Tanaka}, {Tanaka}, {Tanaka}, {Terai}, {Terashima}, {Toba}, {Tominaga}, {Toshikawa}, {Turner}, {Uchida}, {Uchiyama}, {Umetsu}, {Uraguchi}, {Urata}, {Usuda}, {Utsumi}, {Wang}, {Wang}, {Wong}, {Yabe}, {Yamada}, {Yamanoi}, {Yasuda}, {Yeh}, {Yonehara}, \& {Yuma}}]{2018PASJ...70S...4A}
{Aihara}, H., {Arimoto}, N., {Armstrong}, R., {et~al.} 2018, \pasj, 70, S4, \dodoi{10.1093/pasj/psx066}

\bibitem[{{Aihara} {et~al.}(2019){Aihara}, {AlSayyad}, {Ando}, {Armstrong}, {Bosch}, {Egami}, {Furusawa}, {Furusawa}, {Goulding}, {Harikane}, {Hikage}, {Ho}, {Hsieh}, {Huang}, {Ikeda}, {Imanishi}, {Ito}, {Iwata}, {Jaelani}, {Kakuma}, {Kawana}, {Kikuta}, {Kobayashi}, {Koike}, {Komiyama}, {Li}, {Liang}, {Lin}, {Luo}, {Lupton}, {Lust}, {MacArthur}, {Matsuoka}, {Mineo}, {Miyatake}, {Miyazaki}, {More}, {Murata}, {Namiki}, {Nishizawa}, {Oguri}, {Okabe}, {Okamoto}, {Okura}, {Ono}, {Onodera}, {Onoue}, {Osato}, {Ouchi}, {Shibuya}, {Strauss}, {Sugiyama}, {Suto}, {Takada}, {Takagi}, {Takata}, {Takita}, {Tanaka}, {Terai}, {Toba}, {Uchiyama}, {Utsumi}, {Wang}, {Wang}, \& {Yamada}}]{2019PASJ...71..114A}
{Aihara}, H., {AlSayyad}, Y., {Ando}, M., {et~al.} 2019, \pasj, 71, 114, \dodoi{10.1093/pasj/psz103}

\bibitem[{{Alberts} {et~al.}(2014){Alberts}, {Pope}, {Brodwin}, {Atlee}, {Lin}, {Dey}, {Eisenhardt}, {Gettings}, {Gonzalez}, {Jannuzi}, {Mancone}, {Moustakas}, {Snyder}, {Stanford}, {Stern}, {Weiner}, \& {Zeimann}}]{2014MNRAS.437..437A}
{Alberts}, S., {Pope}, A., {Brodwin}, M., {et~al.} 2014, \mnras, 437, 437, \dodoi{10.1093/mnras/stt1897}

\bibitem[{{Alberts} {et~al.}(2016){Alberts}, {Pope}, {Brodwin}, {Chung}, {Cybulski}, {Dey}, {Eisenhardt}, {Galametz}, {Gonzalez}, {Jannuzi}, {Stanford}, {Snyder}, {Stern}, \& {Zeimann}}]{2016ApJ...825...72A}
---. 2016, \apj, 825, 72, \dodoi{10.3847/0004-637X/825/1/72}

\bibitem[{{Balogh} {et~al.}(2004){Balogh}, {Baldry}, {Nichol}, {Miller}, {Bower}, \& {Glazebrook}}]{2004ApJ...615L.101B}
{Balogh}, M.~L., {Baldry}, I.~K., {Nichol}, R., {et~al.} 2004, \apjl, 615, L101, \dodoi{10.1086/426079}

\bibitem[{{Beckmann} {et~al.}(2017){Beckmann}, {Devriendt}, {Slyz}, {Peirani}, {Richardson}, {Dubois}, {Pichon}, {Chisari}, {Kaviraj}, {Laigle}, \& {Volonteri}}]{2017MNRAS.472..949B}
{Beckmann}, R.~S., {Devriendt}, J., {Slyz}, A., {et~al.} 2017, \mnras, 472, 949, \dodoi{10.1093/mnras/stx1831}

\bibitem[{{Boquien} {et~al.}(2019){Boquien}, {Burgarella}, {Roehlly}, {Buat}, {Ciesla}, {Corre}, {Inoue}, \& {Salas}}]{2019A&A...622A.103B}
{Boquien}, M., {Burgarella}, D., {Roehlly}, Y., {et~al.} 2019, \aap, 622, A103, \dodoi{10.1051/0004-6361/201834156}

\bibitem[{{Bruzual} \& {Charlot}(2003)}]{2003MNRAS.344.1000B}
{Bruzual}, G., \& {Charlot}, S. 2003, \mnras, 344, 1000, \dodoi{10.1046/j.1365-8711.2003.06897.x}

\bibitem[{{Burgarella} {et~al.}(2005){Burgarella}, {Buat}, \& {Iglesias-P{\'a}ramo}}]{2005MNRAS.360.1413B}
{Burgarella}, D., {Buat}, V., \& {Iglesias-P{\'a}ramo}, J. 2005, \mnras, 360, 1413, \dodoi{10.1111/j.1365-2966.2005.09131.x}

\bibitem[{{Butcher} \& {Oemler}(1978)}]{1978ApJ...219...18B}
{Butcher}, H., \& {Oemler}, A., J. 1978, \apj, 219, 18, \dodoi{10.1086/155751}

\bibitem[{{Butcher} \& {Oemler}(1984)}]{1984ApJ...285..426B}
---. 1984, \apj, 285, 426, \dodoi{10.1086/162519}

\bibitem[{{Cai} {et~al.}(2017){Cai}, {Fan}, {Bian}, {Zabludoff}, {Yang}, {Prochaska}, {McGreer}, {Zheng}, {Kashikawa}, {Wang}, {Frye}, {Green}, \& {Jiang}}]{2017ApJ...839..131C}
{Cai}, Z., {Fan}, X., {Bian}, F., {et~al.} 2017, \apj, 839, 131, \dodoi{10.3847/1538-4357/aa6a1a}

\bibitem[{{Calzetti} {et~al.}(2000){Calzetti}, {Armus}, {Bohlin}, {Kinney}, {Koornneef}, \& {Storchi-Bergmann}}]{2000ApJ...533..682C}
{Calzetti}, D., {Armus}, L., {Bohlin}, R.~C., {et~al.} 2000, \apj, 533, 682, \dodoi{10.1086/308692}

\bibitem[{{Capak} {et~al.}(2007){Capak}, {Abraham}, {Ellis}, {Mobasher}, {Scoville}, {Sheth}, \& {Koekemoer}}]{2007ApJS..172..284C}
{Capak}, P., {Abraham}, R.~G., {Ellis}, R.~S., {et~al.} 2007, \apjs, 172, 284, \dodoi{10.1086/518424}

\bibitem[{{Casey} {et~al.}(2023){Casey}, {Kartaltepe}, {Drakos}, {Franco}, {Harish}, {Paquereau}, {Ilbert}, {Rose}, {Cox}, {Nightingale}, {Robertson}, {Silverman}, {Koekemoer}, {Massey}, {McCracken}, {Rhodes}, {Akins}, {Allen}, {Amvrosiadis}, {Arango-Toro}, {Bagley}, {Bongiorno}, {Capak}, {Champagne}, {Chartab}, {Ch{\'a}vez Ortiz}, {Chworowsky}, {Cooke}, {Cooper}, {Darvish}, {Ding}, {Faisst}, {Finkelstein}, {Fujimoto}, {Gentile}, {Gillman}, {Gould}, {Gozaliasl}, {Hayward}, {He}, {Hemmati}, {Hirschmann}, {Jahnke}, {Jin}, {Khostovan}, {Kokorev}, {Lambrides}, {Laigle}, {Larson}, {Leung}, {Liu}, {Liaudat}, {Long}, {Magdis}, {Mahler}, {Mainieri}, {Manning}, {Maraston}, {Martin}, {McCleary}, {McKinney}, {McPartland}, {Mobasher}, {Pattnaik}, {Renzini}, {Rich}, {Sanders}, {Sattari}, {Scognamiglio}, {Scoville}, {Sheth}, {Shuntov}, {Sparre}, {Suzuki}, {Talia}, {Toft}, {Trakhtenbrot}, {Urry}, {Valentino}, {Vanderhoof}, {Vardoulaki}, {Weaver}, {Whitaker}, {Wilkins}, {Yang}, \& {Zavala}}]{2023ApJ...954...31C}
{Casey}, C.~M., {Kartaltepe}, J.~S., {Drakos}, N.~E., {et~al.} 2023, \apj, 954, 31, \dodoi{10.3847/1538-4357/acc2bc}

\bibitem[{{Castignani} {et~al.}(2020){Castignani}, {Combes}, \& {Salom{\'e}}}]{2020A&A...635L..10C}
{Castignani}, G., {Combes}, F., \& {Salom{\'e}}, P. 2020, \aap, 635, L10, \dodoi{10.1051/0004-6361/201937155}

\bibitem[{{Chabrier}(2003)}]{Chabrier_2003}
{Chabrier}, G. 2003, \pasp, 115, 763, \dodoi{10.1086/376392}

\bibitem[{{Civano} {et~al.}(2016){Civano}, {Marchesi}, {Comastri}, {Urry}, {Elvis}, {Cappelluti}, {Puccetti}, {Brusa}, {Zamorani}, {Hasinger}, {Aldcroft}, {Alexander}, {Allevato}, {Brunner}, {Capak}, {Finoguenov}, {Fiore}, {Fruscione}, {Gilli}, {Glotfelty}, {Griffiths}, {Hao}, {Harrison}, {Jahnke}, {Kartaltepe}, {Karim}, {LaMassa}, {Lanzuisi}, {Miyaji}, {Ranalli}, {Salvato}, {Sargent}, {Scoville}, {Schawinski}, {Schinnerer}, {Silverman}, {Smolcic}, {Stern}, {Toft}, {Trakhtenbrot}, {Treister}, \& {Vignali}}]{2016ApJ...819...62C}
{Civano}, F., {Marchesi}, S., {Comastri}, A., {et~al.} 2016, \apj, 819, 62, \dodoi{10.3847/0004-637X/819/1/62}

\bibitem[{{Conselice}(2014)}]{2014ARA&A..52..291C}
{Conselice}, C.~J. 2014, \araa, 52, 291, \dodoi{10.1146/annurev-astro-081913-040037}

\bibitem[{{Cooper} {et~al.}(2008){Cooper}, {Newman}, {Weiner}, {Yan}, {Willmer}, {Bundy}, {Coil}, {Conselice}, {Davis}, {Faber}, {Gerke}, {Guhathakurta}, {Koo}, \& {Noeske}}]{2008MNRAS.383.1058C}
{Cooper}, M.~C., {Newman}, J.~A., {Weiner}, B.~J., {et~al.} 2008, \mnras, 383, 1058, \dodoi{10.1111/j.1365-2966.2007.12613.x}

\bibitem[{{Couch} {et~al.}(2001){Couch}, {Balogh}, {Bower}, {Smail}, {Glazebrook}, \& {Taylor}}]{2001ApJ...549..820C}
{Couch}, W.~J., {Balogh}, M.~L., {Bower}, R.~G., {et~al.} 2001, \apj, 549, 820, \dodoi{10.1086/319459}

\bibitem[{{D'Amato} {et~al.}(2020){D'Amato}, {Gilli}, {Prandoni}, {Vignali}, {Massardi}, {Mignoli}, {Cucciati}, {Morishita}, {Decarli}, {Brusa}, {Calura}, {Balmaverde}, {Chiaberge}, {Liuzzo}, {Nanni}, {Peca}, {Pensabene}, {Tozzi}, \& {Norman}}]{2020A&A...641L...6D}
{D'Amato}, Q., {Gilli}, R., {Prandoni}, I., {et~al.} 2020, \aap, 641, L6, \dodoi{10.1051/0004-6361/202038711}

\bibitem[{{Dekel} \& {Birnboim}(2006)}]{2006MNRAS.368....2D}
{Dekel}, A., \& {Birnboim}, Y. 2006, \mnras, 368, 2, \dodoi{10.1111/j.1365-2966.2006.10145.x}

\bibitem[{{Dekel} {et~al.}(2009){Dekel}, {Birnboim}, {Engel}, {Freundlich}, {Goerdt}, {Mumcuoglu}, {Neistein}, {Pichon}, {Teyssier}, \& {Zinger}}]{2009Natur.457..451D}
{Dekel}, A., {Birnboim}, Y., {Engel}, G., {et~al.} 2009, \nat, 457, 451, \dodoi{10.1038/nature07648}

\bibitem[{{Digby-North} {et~al.}(2010){Digby-North}, {Nandra}, {Laird}, {Steidel}, {Georgakakis}, {Bogosavljevi{\'c}}, {Erb}, {Shapley}, {Reddy}, \& {Aird}}]{2010MNRAS.407..846D}
{Digby-North}, J.~A., {Nandra}, K., {Laird}, E.~S., {et~al.} 2010, \mnras, 407, 846, \dodoi{10.1111/j.1365-2966.2010.16977.x}

\bibitem[{{Dressler}(1980)}]{1980ApJ...236..351D}
{Dressler}, A. 1980, \apj, 236, 351, \dodoi{10.1086/157753}

\bibitem[{{Elbaz} {et~al.}(2007){Elbaz}, {Daddi}, {Le Borgne}, {Dickinson}, {Alexander}, {Chary}, {Starck}, {Brandt}, {Kitzbichler}, {MacDonald}, {Nonino}, {Popesso}, {Stern}, \& {Vanzella}}]{2007A&A...468...33E}
{Elbaz}, D., {Daddi}, E., {Le Borgne}, D., {et~al.} 2007, \aap, 468, 33, \dodoi{10.1051/0004-6361:20077525}

\bibitem[{{Ferreira} {et~al.}(2022){Ferreira}, {Adams}, {Conselice}, {Sazonova}, {Austin}, {Caruana}, {Ferrari}, {Verma}, {Trussler}, {Broadhurst}, {Diego}, {Frye}, {Pascale}, {Wilkins}, {Windhorst}, \& {Zitrin}}]{2022ApJ...938L...2F}
{Ferreira}, L., {Adams}, N., {Conselice}, C.~J., {et~al.} 2022, \apjl, 938, L2, \dodoi{10.3847/2041-8213/ac947c}

\bibitem[{{Ferreira} {et~al.}(2023){Ferreira}, {Conselice}, {Sazonova}, {Ferrari}, {Caruana}, {Tohill}, {Lucatelli}, {Adams}, {Irodotou}, {Marshall}, {Roper}, {Lovell}, {Verma}, {Austin}, {Trussler}, \& {Wilkins}}]{2023ApJ...955...94F}
{Ferreira}, L., {Conselice}, C.~J., {Sazonova}, E., {et~al.} 2023, \apj, 955, 94, \dodoi{10.3847/1538-4357/acec76}

\bibitem[{{Genel} {et~al.}(2008){Genel}, {Genzel}, {Bouch{\'e}}, {Sternberg}, {Naab}, {F{\"o}rster Schreiber}, {Shapiro}, {Tacconi}, {Lutz}, {Cresci}, {Buschkamp}, {Davies}, \& {Hicks}}]{2008ApJ...688..789G}
{Genel}, S., {Genzel}, R., {Bouch{\'e}}, N., {et~al.} 2008, \apj, 688, 789, \dodoi{10.1086/592241}

\bibitem[{{Gisler}(1978)}]{1978MNRAS.183..633G}
{Gisler}, G.~R. 1978, \mnras, 183, 633, \dodoi{10.1093/mnras/183.4.633}

\bibitem[{{Goto} {et~al.}(2003){Goto}, {Yamauchi}, {Fujita}, {Okamura}, {Sekiguchi}, {Smail}, {Bernardi}, \& {Gomez}}]{2003MNRAS.346..601G}
{Goto}, T., {Yamauchi}, C., {Fujita}, Y., {et~al.} 2003, \mnras, 346, 601, \dodoi{10.1046/j.1365-2966.2003.07114.x}

\bibitem[{{Hashimoto} {et~al.}(1998){Hashimoto}, {Oemler}, {Lin}, \& {Tucker}}]{1998ApJ...499..589H}
{Hashimoto}, Y., {Oemler}, Augustus, J., {Lin}, H., \& {Tucker}, D.~L. 1998, \apj, 499, 589, \dodoi{10.1086/305657}

\bibitem[{{Hasinger} {et~al.}(2018){Hasinger}, {Capak}, {Salvato}, {Barger}, {Cowie}, {Faisst}, {Hemmati}, {Kakazu}, {Kartaltepe}, {Masters}, {Mobasher}, {Nayyeri}, {Sanders}, {Scoville}, {Suh}, {Steinhardt}, \& {Yang}}]{2018ApJ...858...77H}
{Hasinger}, G., {Capak}, P., {Salvato}, M., {et~al.} 2018, \apj, 858, 77, \dodoi{10.3847/1538-4357/aabacf}

\bibitem[{{Hayashi} {et~al.}(2020){Hayashi}, {Shimakawa}, {Tanaka}, {Onodera}, {Koyama}, {Inoue}, {Komiyama}, {Lee}, {Lin}, \& {Yabe}}]{2020PASJ...72...86H}
{Hayashi}, M., {Shimakawa}, R., {Tanaka}, M., {et~al.} 2020, \pasj, 72, 86, \dodoi{10.1093/pasj/psaa076}

\bibitem[{{Hopkins} {et~al.}(2014){Hopkins}, {Kere{\v{s}}}, {O{\~n}orbe}, {Faucher-Gigu{\`e}re}, {Quataert}, {Murray}, \& {Bullock}}]{2014MNRAS.445..581H}
{Hopkins}, P.~F., {Kere{\v{s}}}, D., {O{\~n}orbe}, J., {et~al.} 2014, \mnras, 445, 581, \dodoi{10.1093/mnras/stu1738}

\bibitem[{{Hoyos} {et~al.}(2016){Hoyos}, {Arag{\'o}n-Salamanca}, {Gray}, {Wolf}, {Maltby}, {Bell}, {B{\"o}hm}, \& {Jogee}}]{2016MNRAS.455..295H}
{Hoyos}, C., {Arag{\'o}n-Salamanca}, A., {Gray}, M.~E., {et~al.} 2016, \mnras, 455, 295, \dodoi{10.1093/mnras/stv2321}

\bibitem[{{Huang} {et~al.}(2022){Huang}, {Lee}, {Cucciati}, {Lemaux}, {Sawicki}, {Malavasi}, {Ramakrishnan}, {Xue}, {Cassara}, {Chiang}, {Dey}, {Gwyn}, {Hathi}, {Pentericci}, {Prescott}, \& {Zamorani}}]{2022ApJ...941..134H}
{Huang}, Y., {Lee}, K.-S., {Cucciati}, O., {et~al.} 2022, \apj, 941, 134, \dodoi{10.3847/1538-4357/ac9ea4}

\bibitem[{{Ilbert} {et~al.}(2013){Ilbert}, {McCracken}, {Le F{\`e}vre}, {Capak}, {Dunlop}, {Karim}, {Renzini}, {Caputi}, {Boissier}, {Arnouts}, {Aussel}, {Comparat}, {Guo}, {Hudelot}, {Kartaltepe}, {Kneib}, {Krogager}, {Le Floc'h}, {Lilly}, {Mellier}, {Milvang-Jensen}, {Moutard}, {Onodera}, {Richard}, {Salvato}, {Sanders}, {Scoville}, {Silverman}, {Taniguchi}, {Tasca}, {Thomas}, {Toft}, {Tresse}, {Vergani}, {Wolk}, \& {Zirm}}]{2013A&A...556A..55I}
{Ilbert}, O., {McCracken}, H.~J., {Le F{\`e}vre}, O., {et~al.} 2013, \aap, 556, A55, \dodoi{10.1051/0004-6361/201321100}

\bibitem[{{Kartaltepe} {et~al.}(2023){Kartaltepe}, {Rose}, {Vanderhoof}, {McGrath}, {Costantin}, {Cox}, {Yung}, {Kocevski}, {Wuyts}, {Ferguson}, {Bagley}, {Finkelstein}, {Amor{\'\i}n}, {Andrews}, {Arrabal Haro}, {Backhaus}, {Behroozi}, {Bisigello}, {Calabr{\`o}}, {Casey}, {Coogan}, {Cooper}, {Croton}, {de la Vega}, {Dickinson}, {Fontana}, {Franco}, {Grazian}, {Grogin}, {Hathi}, {Holwerda}, {Huertas-Company}, {Iyer}, {Jogee}, {Jung}, {Kewley}, {Kirkpatrick}, {Koekemoer}, {Liu}, {Lotz}, {Lucas}, {Newman}, {Pacifici}, {Pandya}, {Papovich}, {Pentericci}, {P{\'e}rez-Gonz{\'a}lez}, {Petersen}, {Pirzkal}, {Rafelski}, {Ravindranath}, {Simons}, {Snyder}, {Somerville}, {Stanway}, {Straughn}, {Tacchella}, {Trump}, {Vega-Ferrero}, {Wilkins}, {Yang}, \& {Zavala}}]{2023ApJ...946L..15K}
{Kartaltepe}, J.~S., {Rose}, C., {Vanderhoof}, B.~N., {et~al.} 2023, \apjl, 946, L15, \dodoi{10.3847/2041-8213/acad01}

\bibitem[{{Kashino} {et~al.}(2013){Kashino}, {Silverman}, {Rodighiero}, {Renzini}, {Arimoto}, {Daddi}, {Lilly}, {Sanders}, {Kartaltepe}, {Zahid}, {Nagao}, {Sugiyama}, {Capak}, {Carollo}, {Chu}, {Hasinger}, {Ilbert}, {Kajisawa}, {Kewley}, {Koekemoer}, {Kova{\v{c}}}, {Le F{\`e}vre}, {Masters}, {McCracken}, {Onodera}, {Scoville}, {Strazzullo}, {Symeonidis}, \& {Taniguchi}}]{2013ApJ...777L...8K}
{Kashino}, D., {Silverman}, J.~D., {Rodighiero}, G., {et~al.} 2013, \apjl, 777, L8, \dodoi{10.1088/2041-8205/777/1/L8}

\bibitem[{{Kashino} {et~al.}(2017){Kashino}, {Silverman}, {Sanders}, {Kartaltepe}, {Daddi}, {Renzini}, {Valentino}, {Rodighiero}, {Juneau}, {Kewley}, {Zahid}, {Arimoto}, {Nagao}, {Chu}, {Sugiyama}, {Civano}, {Ilbert}, {Kajisawa}, {Le F{\`e}vre}, {Maier}, {Masters}, {Miyaji}, {Onodera}, {Puglisi}, \& {Taniguchi}}]{2017ApJ...835...88K}
{Kashino}, D., {Silverman}, J.~D., {Sanders}, D., {et~al.} 2017, \apj, 835, 88, \dodoi{10.3847/1538-4357/835/1/88}

\bibitem[{{Kauffmann} {et~al.}(2004){Kauffmann}, {White}, {Heckman}, {M{\'e}nard}, {Brinchmann}, {Charlot}, {Tremonti}, \& {Brinkmann}}]{2004MNRAS.353..713K}
{Kauffmann}, G., {White}, S. D.~M., {Heckman}, T.~M., {et~al.} 2004, \mnras, 353, 713, \dodoi{10.1111/j.1365-2966.2004.08117.x}

\bibitem[{{Kennicutt}(1998)}]{1998ARA&A..36..189K}
{Kennicutt}, Robert~C., J. 1998, \araa, 36, 189, \dodoi{10.1146/annurev.astro.36.1.189}

\bibitem[{{Kere{\v{s}}} {et~al.}(2005){Kere{\v{s}}}, {Katz}, {Weinberg}, \& {Dav{\'e}}}]{2005MNRAS.363....2K}
{Kere{\v{s}}}, D., {Katz}, N., {Weinberg}, D.~H., \& {Dav{\'e}}, R. 2005, \mnras, 363, 2, \dodoi{10.1111/j.1365-2966.2005.09451.x}

\bibitem[{{Kodama} \& {Smail}(2001)}]{2001MNRAS.326..637K}
{Kodama}, T., \& {Smail}, I. 2001, \mnras, 326, 637, \dodoi{10.1046/j.1365-8711.2001.04624.x}

\bibitem[{{Kodama} {et~al.}(2001){Kodama}, {Smail}, {Nakata}, {Okamura}, \& {Bower}}]{2001ApJ...562L...9K}
{Kodama}, T., {Smail}, I., {Nakata}, F., {Okamura}, S., \& {Bower}, R.~G. 2001, \apjl, 562, L9, \dodoi{10.1086/338100}

\bibitem[{{Koyama} {et~al.}(2019){Koyama}, {Shimakawa}, {Yamamura}, {Kodama}, \& {Hayashi}}]{2019PASJ...71....8K}
{Koyama}, Y., {Shimakawa}, R., {Yamamura}, I., {Kodama}, T., \& {Hayashi}, M. 2019, \pasj, 71, 8, \dodoi{10.1093/pasj/psy113}

\bibitem[{{Kriek} {et~al.}(2015){Kriek}, {Shapley}, {Reddy}, {Siana}, {Coil}, {Mobasher}, {Freeman}, {de Groot}, {Price}, {Sanders}, {Shivaei}, {Brammer}, {Momcheva}, {Skelton}, {van Dokkum}, {Whitaker}, {Aird}, {Azadi}, {Kassis}, {Bullock}, {Conroy}, {Dav{\'e}}, {Kere{\v{s}}}, \& {Krumholz}}]{2015ApJS..218...15K}
{Kriek}, M., {Shapley}, A.~E., {Reddy}, N.~A., {et~al.} 2015, \apjs, 218, 15, \dodoi{10.1088/0067-0049/218/2/15}

\bibitem[{{Krishnan} {et~al.}(2017){Krishnan}, {Hatch}, {Almaini}, {Kocevski}, {Cooke}, {Hartley}, {Hasinger}, {Maltby}, {Muldrew}, \& {Simpson}}]{2017MNRAS.470.2170K}
{Krishnan}, C., {Hatch}, N.~A., {Almaini}, O., {et~al.} 2017, \mnras, 470, 2170, \dodoi{10.1093/mnras/stx1315}

\bibitem[{{Lee} {et~al.}(2016){Lee}, {Hennawi}, {White}, {Prochaska}, {Font-Ribera}, {Schlegel}, {Rich}, {Suzuki}, {Stark}, {Le F{\`e}vre}, {Nugent}, {Salvato}, \& {Zamorani}}]{2016ApJ...817..160L}
{Lee}, K.-G., {Hennawi}, J.~F., {White}, M., {et~al.} 2016, \apj, 817, 160, \dodoi{10.3847/0004-637X/817/2/160}

\bibitem[{{Lehmer} {et~al.}(2009){Lehmer}, {Alexander}, {Geach}, {Smail}, {Basu-Zych}, {Bauer}, {Chapman}, {Matsuda}, {Scharf}, {Volonteri}, \& {Yamada}}]{2009ApJ...691..687L}
{Lehmer}, B.~D., {Alexander}, D.~M., {Geach}, J.~E., {et~al.} 2009, \apj, 691, 687, \dodoi{10.1088/0004-637X/691/1/687}

\bibitem[{{Lemaux} {et~al.}(2022){Lemaux}, {Cucciati}, {Le F{\`e}vre}, {Zamorani}, {Lubin}, {Hathi}, {Ilbert}, {Pelliccia}, {Amor{\'\i}n}, {Bardelli}, {Cassata}, {Gal}, {Garilli}, {Guaita}, {Giavalisco}, {Hung}, {Koekemoer}, {Maccagni}, {Pentericci}, {Ribeiro}, {Schaerer}, {Shah}, {Shen}, {Staab}, {Talia}, {Thomas}, {Tomczak}, {Tresse}, {Vanzella}, {Vergani}, \& {Zucca}}]{2022A&A...662A..33L}
{Lemaux}, B.~C., {Cucciati}, O., {Le F{\`e}vre}, O., {et~al.} 2022, \aap, 662, A33, \dodoi{10.1051/0004-6361/202039346}

\bibitem[{{Lotz} {et~al.}(2010){Lotz}, {Jonsson}, {Cox}, \& {Primack}}]{2010MNRAS.404..575L}
{Lotz}, J.~M., {Jonsson}, P., {Cox}, T.~J., \& {Primack}, J.~R. 2010, \mnras, 404, 575, \dodoi{10.1111/j.1365-2966.2010.16268.x}

\bibitem[{{Lotz} {et~al.}(2004){Lotz}, {Primack}, \& {Madau}}]{2004AJ....128..163L}
{Lotz}, J.~M., {Primack}, J., \& {Madau}, P. 2004, \aj, 128, 163, \dodoi{10.1086/421849}

\bibitem[{{Lotz} {et~al.}(2008){Lotz}, {Davis}, {Faber}, {Guhathakurta}, {Gwyn}, {Huang}, {Koo}, {Le Floc'h}, {Lin}, {Newman}, {Noeske}, {Papovich}, {Willmer}, {Coil}, {Conselice}, {Cooper}, {Hopkins}, {Metevier}, {Primack}, {Rieke}, \& {Weiner}}]{2008ApJ...672..177L}
{Lotz}, J.~M., {Davis}, M., {Faber}, S.~M., {et~al.} 2008, \apj, 672, 177, \dodoi{10.1086/523659}

\bibitem[{{Marchesi} {et~al.}(2016){Marchesi}, {Civano}, {Elvis}, {Salvato}, {Brusa}, {Comastri}, {Gilli}, {Hasinger}, {Lanzuisi}, {Miyaji}, {Treister}, {Urry}, {Vignali}, {Zamorani}, {Allevato}, {Cappelluti}, {Cardamone}, {Finoguenov}, {Griffiths}, {Karim}, {Laigle}, {LaMassa}, {Jahnke}, {Ranalli}, {Schawinski}, {Schinnerer}, {Silverman}, {Smolcic}, {Suh}, \& {Trakhtenbrot}}]{2016ApJ...817...34M}
{Marchesi}, S., {Civano}, F., {Elvis}, M., {et~al.} 2016, \apj, 817, 34, \dodoi{10.3847/0004-637X/817/1/34}

\bibitem[{{Mawatari} {et~al.}(2017){Mawatari}, {Inoue}, {Yamada}, {Hayashino}, {Otsuka}, {Matsuda}, {Umehata}, {Ouchi}, \& {Mukae}}]{2017MNRAS.467.3951M}
{Mawatari}, K., {Inoue}, A.~K., {Yamada}, T., {et~al.} 2017, \mnras, 467, 3951, \dodoi{10.1093/mnras/stx038}

\bibitem[{{Mei} {et~al.}(2023){Mei}, {Hatch}, {Amodeo}, {Afanasiev}, {De Breuck}, {Stern}, {Cooke}, {Gonzalez}, {Noirot}, {Rettura}, {Seymour}, {Stanford}, {Vernet}, \& {Wylezalek}}]{2023A&A...670A..58M}
{Mei}, S., {Hatch}, N.~A., {Amodeo}, S., {et~al.} 2023, \aap, 670, A58, \dodoi{10.1051/0004-6361/202243551}

\bibitem[{{Mo} {et~al.}(2018){Mo}, {Gonzalez}, {Stern}, {Brodwin}, {Decker}, {Eisenhardt}, {Moravec}, {Stanford}, \& {Wylezalek}}]{2018ApJ...869..131M}
{Mo}, W., {Gonzalez}, A., {Stern}, D., {et~al.} 2018, \apj, 869, 131, \dodoi{10.3847/1538-4357/aaef83}

\bibitem[{{Monson} {et~al.}(2021){Monson}, {Lehmer}, {Doore}, {Eufrasio}, {Bonine}, {Alexander}, {Harrison}, {Kubo}, {Mantha}, {Saez}, {Straughn}, \& {Umehata}}]{2021ApJ...919...51M}
{Monson}, E.~B., {Lehmer}, B.~D., {Doore}, K., {et~al.} 2021, \apj, 919, 51, \dodoi{10.3847/1538-4357/ac0f84}

\bibitem[{{Morishita} {et~al.}(2017){Morishita}, {Abramson}, {Treu}, {Vulcani}, {Schmidt}, {Dressler}, {Poggianti}, {Malkan}, {Wang}, {Huang}, {Trenti}, {Brada{\v{c}}}, \& {Hoag}}]{morishita17}
{Morishita}, T., {Abramson}, L.~E., {Treu}, T., {et~al.} 2017, \apj, 835, 254, \dodoi{10.3847/1538-4357/835/2/254}

\bibitem[{{Morishita} {et~al.}(2023){Morishita}, {Stiavelli}, {Chary}, {Trenti}, {Bergamini}, {Chiaberge}, {Leethochawalit}, {Roberts-Borsani}, {Shen}, \& {Treu}}]{morishita23}
{Morishita}, T., {Stiavelli}, M., {Chary}, R.-R., {et~al.} 2023, arXiv e-prints, arXiv:2308.05018, \dodoi{10.48550/arXiv.2308.05018}

\bibitem[{{Naufal} {et~al.}(2023){Naufal}, {Koyama}, {Shimakawa}, \& {Kodama}}]{2023ApJ...958..170N}
{Naufal}, A., {Koyama}, Y., {Shimakawa}, R., \& {Kodama}, T. 2023, \apj, 958, 170, \dodoi{10.3847/1538-4357/acfb81}

\bibitem[{{Nelson} {et~al.}(2019){Nelson}, {Pillepich}, {Springel}, {Pakmor}, {Weinberger}, {Genel}, {Torrey}, {Vogelsberger}, {Marinacci}, \& {Hernquist}}]{2019MNRAS.490.3234N}
{Nelson}, D., {Pillepich}, A., {Springel}, V., {et~al.} 2019, \mnras, 490, 3234, \dodoi{10.1093/mnras/stz2306}

\bibitem[{{Newman} {et~al.}(2014){Newman}, {Ellis}, {Andreon}, {Treu}, {Raichoor}, \& {Trinchieri}}]{2014ApJ...788...51N}
{Newman}, A.~B., {Ellis}, R.~S., {Andreon}, S., {et~al.} 2014, \apj, 788, 51, \dodoi{10.1088/0004-637X/788/1/51}

\bibitem[{{Noll} {et~al.}(2009){Noll}, {Burgarella}, {Giovannoli}, {Buat}, {Marcillac}, \& {Mu{\~n}oz-Mateos}}]{2009A&A...507.1793N}
{Noll}, S., {Burgarella}, D., {Giovannoli}, E., {et~al.} 2009, \aap, 507, 1793, \dodoi{10.1051/0004-6361/200912497}

\bibitem[{{Oke} \& {Gunn}(1983)}]{1983ApJ...266..713O}
{Oke}, J.~B., \& {Gunn}, J.~E. 1983, \apj, 266, 713, \dodoi{10.1086/160817}

\bibitem[{{Papovich} {et~al.}(2012){Papovich}, {Bassett}, {Lotz}, {van der Wel}, {Tran}, {Finkelstein}, {Bell}, {Conselice}, {Dekel}, {Dunlop}, {Guo}, {Faber}, {Farrah}, {Ferguson}, {Finkelstein}, {H{\"a}ussler}, {Kocevski}, {Koekemoer}, {Koo}, {McGrath}, {McLure}, {McIntosh}, {Momcheva}, {Newman}, {Rudnick}, {Weiner}, {Willmer}, \& {Wuyts}}]{2012ApJ...750...93P}
{Papovich}, C., {Bassett}, R., {Lotz}, J.~M., {et~al.} 2012, \apj, 750, 93, \dodoi{10.1088/0004-637X/750/2/93}

\bibitem[{{Paulino-Afonso} {et~al.}(2019){Paulino-Afonso}, {Sobral}, {Darvish}, {Ribeiro}, {van der Wel}, {Stott}, {Buitrago}, {Best}, {Stroe}, \& {Craig}}]{2019A&A...630A..57P}
{Paulino-Afonso}, A., {Sobral}, D., {Darvish}, B., {et~al.} 2019, \aap, 630, A57, \dodoi{10.1051/0004-6361/201935137}

\bibitem[{{Peng} {et~al.}(2002){Peng}, {Ho}, {Impey}, \& {Rix}}]{2002AJ....124..266P}
{Peng}, C.~Y., {Ho}, L.~C., {Impey}, C.~D., \& {Rix}, H.-W. 2002, \aj, 124, 266, \dodoi{10.1086/340952}

\bibitem[{{Peng} {et~al.}(2010){Peng}, {Lilly}, {Kova{\v{c}}}, {Bolzonella}, {Pozzetti}, {Renzini}, {Zamorani}, {Ilbert}, {Knobel}, {Iovino}, {Maier}, {Cucciati}, {Tasca}, {Carollo}, {Silverman}, {Kampczyk}, {de Ravel}, {Sanders}, {Scoville}, {Contini}, {Mainieri}, {Scodeggio}, {Kneib}, {Le F{\`e}vre}, {Bardelli}, {Bongiorno}, {Caputi}, {Coppa}, {de la Torre}, {Franzetti}, {Garilli}, {Lamareille}, {Le Borgne}, {Le Brun}, {Mignoli}, {Perez Montero}, {Pello}, {Ricciardelli}, {Tanaka}, {Tresse}, {Vergani}, {Welikala}, {Zucca}, {Oesch}, {Abbas}, {Barnes}, {Bordoloi}, {Bottini}, {Cappi}, {Cassata}, {Cimatti}, {Fumana}, {Hasinger}, {Koekemoer}, {Leauthaud}, {Maccagni}, {Marinoni}, {McCracken}, {Memeo}, {Meneux}, {Nair}, {Porciani}, {Presotto}, \& {Scaramella}}]{2010ApJ...721..193P}
{Peng}, Y.-j., {Lilly}, S.~J., {Kova{\v{c}}}, K., {et~al.} 2010, \apj, 721, 193, \dodoi{10.1088/0004-637X/721/1/193}

\bibitem[{{Penny} {et~al.}(2018){Penny}, {Masters}, {Smethurst}, {Nichol}, {Krawczyk}, {Bizyaev}, {Greene}, {Liu}, {Marinelli}, {Rembold}, {Riffel}, {Ilha}, {Wylezalek}, {Andrews}, {Bundy}, {Drory}, {Oravetz}, \& {Pan}}]{2018MNRAS.476..979P}
{Penny}, S.~J., {Masters}, K.~L., {Smethurst}, R., {et~al.} 2018, \mnras, 476, 979, \dodoi{10.1093/mnras/sty202}

\bibitem[{{P{\'e}rez-Mart{\'\i}nez} {et~al.}(2024){P{\'e}rez-Mart{\'\i}nez}, {Kodama}, {Koyama}, {Shimakawa}, {Suzuki}, {Daikuhara}, {Adachi}, {Onodera}, \& {Tanaka}}]{2024MNRAS.52710221P}
{P{\'e}rez-Mart{\'\i}nez}, J.~M., {Kodama}, T., {Koyama}, Y., {et~al.} 2024, \mnras, 527, 10221, \dodoi{10.1093/mnras/stad3805}

\bibitem[{{Perrin} {et~al.}(2014){Perrin}, {Sivaramakrishnan}, {Lajoie}, {Elliott}, {Pueyo}, {Ravindranath}, \& {Albert}}]{2014SPIE.9143E..3XP}
{Perrin}, M.~D., {Sivaramakrishnan}, A., {Lajoie}, C.-P., {et~al.} 2014, in Society of Photo-Optical Instrumentation Engineers (SPIE) Conference Series, Vol. 9143, Space Telescopes and Instrumentation 2014: Optical, Infrared, and Millimeter Wave, ed. J.~{Oschmann}, Jacobus~M., M.~{Clampin}, G.~G. {Fazio}, \& H.~A. {MacEwen}, 91433X, \dodoi{10.1117/12.2056689}

\bibitem[{{Peter} {et~al.}(2007){Peter}, {Shapley}, {Law}, {Steidel}, {Erb}, {Reddy}, \& {Pettini}}]{2007ApJ...668...23P}
{Peter}, A. H.~G., {Shapley}, A.~E., {Law}, D.~R., {et~al.} 2007, \apj, 668, 23, \dodoi{10.1086/521184}

\bibitem[{{Peth} {et~al.}(2016){Peth}, {Lotz}, {Freeman}, {McPartland}, {Mortazavi}, {Snyder}, {Barro}, {Grogin}, {Guo}, {Hemmati}, {Kartaltepe}, {Kocevski}, {Koekemoer}, {McIntosh}, {Nayyeri}, {Papovich}, {Primack}, \& {Simons}}]{2016MNRAS.458..963P}
{Peth}, M.~A., {Lotz}, J.~M., {Freeman}, P.~E., {et~al.} 2016, \mnras, 458, 963, \dodoi{10.1093/mnras/stw252}

\bibitem[{{Poggianti} {et~al.}(1999){Poggianti}, {Smail}, {Dressler}, {Couch}, {Barger}, {Butcher}, {Ellis}, \& {Oemler}}]{1999ApJ...518..576P}
{Poggianti}, B.~M., {Smail}, I., {Dressler}, A., {et~al.} 1999, \apj, 518, 576, \dodoi{10.1086/307322}

\bibitem[{{Pozzetti} {et~al.}(2007){Pozzetti}, {Bolzonella}, {Lamareille}, {Zamorani}, {Franzetti}, {Le F{\`e}vre}, {Iovino}, {Temporin}, {Ilbert}, {Arnouts}, {Charlot}, {Brinchmann}, {Zucca}, {Tresse}, {Scodeggio}, {Guzzo}, {Bottini}, {Garilli}, {Le Brun}, {Maccagni}, {Picat}, {Scaramella}, {Vettolani}, {Zanichelli}, {Adami}, {Bardelli}, {Cappi}, {Ciliegi}, {Contini}, {Foucaud}, {Gavignaud}, {McCracken}, {Marano}, {Marinoni}, {Mazure}, {Meneux}, {Merighi}, {Paltani}, {Pell{\`o}}, {Pollo}, {Radovich}, {Bondi}, {Bongiorno}, {Cucciati}, {de la Torre}, {Gregorini}, {Mellier}, {Merluzzi}, {Vergani}, \& {Walcher}}]{Pozzetti_2007}
{Pozzetti}, L., {Bolzonella}, M., {Lamareille}, F., {et~al.} 2007, \aap, 474, 443, \dodoi{10.1051/0004-6361:20077609}

\bibitem[{{Pozzetti} {et~al.}(2010){Pozzetti}, {Bolzonella}, {Zucca}, {Zamorani}, {Lilly}, {Renzini}, {Moresco}, {Mignoli}, {Cassata}, {Tasca}, {Lamareille}, {Maier}, {Meneux}, {Halliday}, {Oesch}, {Vergani}, {Caputi}, {Kova{\v{c}}}, {Cimatti}, {Cucciati}, {Iovino}, {Peng}, {Carollo}, {Contini}, {Kneib}, {Le F{\'e}vre}, {Mainieri}, {Scodeggio}, {Bardelli}, {Bongiorno}, {Coppa}, {de la Torre}, {de Ravel}, {Franzetti}, {Garilli}, {Kampczyk}, {Knobel}, {Le Borgne}, {Le Brun}, {Pell{\`o}}, {Perez Montero}, {Ricciardelli}, {Silverman}, {Tanaka}, {Tresse}, {Abbas}, {Bottini}, {Cappi}, {Guzzo}, {Koekemoer}, {Leauthaud}, {Maccagni}, {Marinoni}, {McCracken}, {Memeo}, {Porciani}, {Scaramella}, {Scarlata}, \& {Scoville}}]{2010A&A...523A..13P}
{Pozzetti}, L., {Bolzonella}, M., {Zucca}, E., {et~al.} 2010, \aap, 523, A13, \dodoi{10.1051/0004-6361/200913020}

\bibitem[{{Price} {et~al.}(2014){Price}, {Kriek}, {Brammer}, {Conroy}, {F{\"o}rster Schreiber}, {Franx}, {Fumagalli}, {Lundgren}, {Momcheva}, {Nelson}, {Skelton}, {van Dokkum}, {Whitaker}, \& {Wuyts}}]{2014ApJ...788...86P}
{Price}, S.~H., {Kriek}, M., {Brammer}, G.~B., {et~al.} 2014, \apj, 788, 86, \dodoi{10.1088/0004-637X/788/1/86}

\bibitem[{{Reddy} {et~al.}(2020){Reddy}, {Shapley}, {Kriek}, {Steidel}, {Shivaei}, {Sanders}, {Mobasher}, {Coil}, {Siana}, {Freeman}, {Azadi}, {Fetherolf}, {Leung}, {Price}, \& {Zick}}]{2020ApJ...902..123R}
{Reddy}, N.~A., {Shapley}, A.~E., {Kriek}, M., {et~al.} 2020, \apj, 902, 123, \dodoi{10.3847/1538-4357/abb674}

\bibitem[{{Rodriguez-Gomez} {et~al.}(2019){Rodriguez-Gomez}, {Snyder}, {Lotz}, {Nelson}, {Pillepich}, {Springel}, {Genel}, {Weinberger}, {Tacchella}, {Pakmor}, {Torrey}, {Marinacci}, {Vogelsberger}, {Hernquist}, \& {Thilker}}]{2019MNRAS.483.4140R}
{Rodriguez-Gomez}, V., {Snyder}, G.~F., {Lotz}, J.~M., {et~al.} 2019, \mnras, 483, 4140, \dodoi{10.1093/mnras/sty3345}

\bibitem[{{Roper} {et~al.}(2022){Roper}, {Lovell}, {Vijayan}, {Marshall}, {Irodotou}, {Kuusisto}, {Thomas}, \& {Wilkins}}]{2022MNRAS.514.1921R}
{Roper}, W.~J., {Lovell}, C.~C., {Vijayan}, A.~P., {et~al.} 2022, \mnras, 514, 1921, \dodoi{10.1093/mnras/stac1368}

\bibitem[{{Salpeter}(1955)}]{1955ApJ...121..161S}
{Salpeter}, E.~E. 1955, \apj, 121, 161, \dodoi{10.1086/145971}

\bibitem[{{Sazonova} {et~al.}(2020){Sazonova}, {Alatalo}, {Lotz}, {Rowlands}, {Snyder}, {Boone}, {Brodwin}, {Hayden}, {Lanz}, {Perlmutter}, \& {Rodriguez-Gomez}}]{2020ApJ...899...85S}
{Sazonova}, E., {Alatalo}, K., {Lotz}, J., {et~al.} 2020, \apj, 899, 85, \dodoi{10.3847/1538-4357/aba42f}

\bibitem[{{Schaye} {et~al.}(2015){Schaye}, {Crain}, {Bower}, {Furlong}, {Schaller}, {Theuns}, {Dalla Vecchia}, {Frenk}, {McCarthy}, {Helly}, {Jenkins}, {Rosas-Guevara}, {White}, {Baes}, {Booth}, {Camps}, {Navarro}, {Qu}, {Rahmati}, {Sawala}, {Thomas}, \& {Trayford}}]{2015MNRAS.446..521S}
{Schaye}, J., {Crain}, R.~A., {Bower}, R.~G., {et~al.} 2015, \mnras, 446, 521, \dodoi{10.1093/mnras/stu2058}

\bibitem[{{Shi} {et~al.}(2024){Shi}, {Malavasi}, {Toshikawa}, \& {Zheng}}]{2024ApJ...961...39S}
{Shi}, K., {Malavasi}, N., {Toshikawa}, J., \& {Zheng}, X. 2024, \apj, 961, 39, \dodoi{10.3847/1538-4357/ad11d7}

\bibitem[{{Shimakawa} {et~al.}(2018){Shimakawa}, {Kodama}, {Hayashi}, {Prochaska}, {Tanaka}, {Cai}, {Suzuki}, {Tadaki}, \& {Koyama}}]{2018MNRAS.473.1977S}
{Shimakawa}, R., {Kodama}, T., {Hayashi}, M., {et~al.} 2018, \mnras, 473, 1977, \dodoi{10.1093/mnras/stx2494}

\bibitem[{{Silverman} {et~al.}(2015){Silverman}, {Kashino}, {Sanders}, {Kartaltepe}, {Arimoto}, {Renzini}, {Rodighiero}, {Daddi}, {Zahid}, {Nagao}, {Kewley}, {Lilly}, {Sugiyama}, {Baronchelli}, {Capak}, {Carollo}, {Chu}, {Hasinger}, {Ilbert}, {Juneau}, {Kajisawa}, {Koekemoer}, {Kovac}, {Le F{\`e}vre}, {Masters}, {McCracken}, {Onodera}, {Schulze}, {Scoville}, {Strazzullo}, \& {Taniguchi}}]{2015ApJS..220...12S}
{Silverman}, J.~D., {Kashino}, D., {Sanders}, D., {et~al.} 2015, \apjs, 220, 12, \dodoi{10.1088/0067-0049/220/1/12}

\bibitem[{{Smith} {et~al.}(2010){Smith}, {Lucey}, {Hammer}, {Hornschemeier}, {Carter}, {Hudson}, {Marzke}, {Mouhcine}, {Eftekharzadeh}, {James}, {Khosroshahi}, {Kourkchi}, \& {Karick}}]{2010MNRAS.408.1417S}
{Smith}, R.~J., {Lucey}, J.~R., {Hammer}, D., {et~al.} 2010, \mnras, 408, 1417, \dodoi{10.1111/j.1365-2966.2010.17253.x}

\bibitem[{{Snyder} {et~al.}(2015{\natexlab{a}}){Snyder}, {Lotz}, {Moody}, {Peth}, {Freeman}, {Ceverino}, {Primack}, \& {Dekel}}]{2015MNRAS.451.4290S}
{Snyder}, G.~F., {Lotz}, J., {Moody}, C., {et~al.} 2015{\natexlab{a}}, \mnras, 451, 4290, \dodoi{10.1093/mnras/stv1231}

\bibitem[{{Snyder} {et~al.}(2015{\natexlab{b}}){Snyder}, {Torrey}, {Lotz}, {Genel}, {McBride}, {Vogelsberger}, {Pillepich}, {Nelson}, {Sales}, {Sijacki}, {Hernquist}, \& {Springel}}]{2015MNRAS.454.1886S}
{Snyder}, G.~F., {Torrey}, P., {Lotz}, J.~M., {et~al.} 2015{\natexlab{b}}, \mnras, 454, 1886, \dodoi{10.1093/mnras/stv2078}

\bibitem[{{Somerville} \& {Dav{\'e}}(2015)}]{2015ARA&A..53...51S}
{Somerville}, R.~S., \& {Dav{\'e}}, R. 2015, \araa, 53, 51, \dodoi{10.1146/annurev-astro-082812-140951}

\bibitem[{{Tasca} {et~al.}(2009){Tasca}, {Kneib}, {Iovino}, {Le F{\`e}vre}, {Kova{\v{c}}}, {Bolzonella}, {Lilly}, {Abraham}, {Cassata}, {Cucciati}, {Guzzo}, {Tresse}, {Zamorani}, {Capak}, {Garilli}, {Scodeggio}, {Sheth}, {Zucca}, {Carollo}, {Contini}, {Mainieri}, {Renzini}, {Bardelli}, {Bongiorno}, {Caputi}, {Coppa}, {de La Torre}, {de Ravel}, {Franzetti}, {Kampczyk}, {Knobel}, {Koekemoer}, {Lamareille}, {Le Borgne}, {Le Brun}, {Maier}, {Mignoli}, {Pello}, {Peng}, {Perez Montero}, {Ricciardelli}, {Silverman}, {Vergani}, {Tanaka}, {Abbas}, {Bottini}, {Cappi}, {Cimatti}, {Ilbert}, {Leauthaud}, {Maccagni}, {Marinoni}, {McCracken}, {Memeo}, {Meneux}, {Oesch}, {Porciani}, {Pozzetti}, {Scaramella}, \& {Scarlata}}]{2009A&A...503..379T}
{Tasca}, L.~A.~M., {Kneib}, J.~P., {Iovino}, A., {et~al.} 2009, \aap, 503, 379, \dodoi{10.1051/0004-6361/200912213}

\bibitem[{{von der Linden} {et~al.}(2010){von der Linden}, {Wild}, {Kauffmann}, {White}, \& {Weinmann}}]{2010MNRAS.404.1231V}
{von der Linden}, A., {Wild}, V., {Kauffmann}, G., {White}, S. D.~M., \& {Weinmann}, S. 2010, \mnras, 404, 1231, \dodoi{10.1111/j.1365-2966.2010.16375.x}

\bibitem[{{Weaver} {et~al.}(2022){Weaver}, {Kauffmann}, {Ilbert}, {McCracken}, {Moneti}, {Toft}, {Brammer}, {Shuntov}, {Davidzon}, {Hsieh}, {Laigle}, {Anastasiou}, {Jespersen}, {Vinther}, {Capak}, {Casey}, {McPartland}, {Milvang-Jensen}, {Mobasher}, {Sanders}, {Zalesky}, {Arnouts}, {Aussel}, {Dunlop}, {Faisst}, {Franx}, {Furtak}, {Fynbo}, {Gould}, {Greve}, {Gwyn}, {Kartaltepe}, {Kashino}, {Koekemoer}, {Kokorev}, {Le F{\`e}vre}, {Lilly}, {Masters}, {Magdis}, {Mehta}, {Peng}, {Riechers}, {Salvato}, {Sawicki}, {Scarlata}, {Scoville}, {Shirley}, {Silverman}, {Sneppen}, {Smolc̆i{\'c}}, {Steinhardt}, {Stern}, {Tanaka}, {Taniguchi}, {Teplitz}, {Vaccari}, {Wang}, \& {Zamorani}}]{2022ApJS..258...11W}
{Weaver}, J.~R., {Kauffmann}, O.~B., {Ilbert}, O., {et~al.} 2022, \apjs, 258, 11, \dodoi{10.3847/1538-4365/ac3078}

\end{thebibliography}
\bibliographystyle{aasjournal}

\end{document}